\documentclass[%
 reprint,
superscriptaddress,
 amsmath,amssymb,
 aps, prl
floatfix,
]{revtex4-2}

\usepackage{graphicx}% Include figure files
\usepackage[caption=false]{subfig}
\usepackage{dcolumn}% Align table columns on decimal point
\usepackage{bm}% bold math
\usepackage{siunitx}
\usepackage{physics}
\usepackage{placeins}
\usepackage{todonotes}
\usepackage{gensymb}
\usepackage{placeins}
\usepackage{color,soul}

\newcommand{\snr}{\text{SNR}}

\newcommand{\Ap}{A^{\prime}}
\newcommand{\betaterm}{\frac{\beta}{\beta+1}}
\newcommand{\gagg}{g_{a\gamma \gamma}}
\newcommand{\veff}{V_{\text{eff}}}

\newcommand{\qdm}{Q_{\text{DM}}}
\newcommand{\ql}{Q_{\text{L}}}
\newcommand{\dfdm}{\Delta f_{\text{DM}}}
\newcommand{\tadd}{T_\text{add}}
\newcommand{\tvts}{T_\text{VTS}}

\newcommand{\tcav}{T_\text{cav}}

\begin{document}

\preprint{APS/123-QED}

\title{Deepest sensitivity to wavelike dark photon dark matter with superconducting radio frequency cavities}% Force line breaks with \\

\author{R. Cervantes}%
\email[Correspondence to: ]{raphaelc@fnal.gov}
\affiliation{Fermi National Accelerator Laboratory, Batavia IL 60510}

\author{J. Aumentado}
\affiliation{National Institute of Standards and Technology, Boulder, CO 80305}

\author{C. Braggio}
\affiliation{Dip. di Fisica e Astronomia, Universit\`{a} di Padova, 35100 Padova, Italy}
\affiliation{INFN - Sezione di Padova, 35100 Padova, Italy}

\author{B. Giaccone}
\affiliation{Fermi National Accelerator Laboratory, Batavia IL 60510}

\author{D. Frolov}
\affiliation{Fermi National Accelerator Laboratory, Batavia IL 60510}

\author{A. Grassellino}
\affiliation{Fermi National Accelerator Laboratory, Batavia IL 60510}

\author{R. Harnik}
\affiliation{Fermi National Accelerator Laboratory, Batavia IL 60510}

\author{F. Lecocq}
\affiliation{National Institute of Standards and Technology, Boulder, CO 80305}

\author{O. Melnychuk}
\affiliation{Fermi National Accelerator Laboratory, Batavia IL 60510}

\author{R. Pilipenko}
\affiliation{Fermi National Accelerator Laboratory, Batavia IL 60510}

\author{S. Posen}
\affiliation{Fermi National Accelerator Laboratory, Batavia IL 60510}

\author{A. Romanenko}
\affiliation{Fermi National Accelerator Laboratory, Batavia IL 60510}

\date{\today}% It is always \today, today,

\begin{abstract}
    Wavelike, bosonic dark matter candidates like axions and dark photons can be detected using microwave cavities known as haloscopes. Traditionally, haloscopes consist of tunable copper cavities operating in the TM$_{010}$ mode, but ohmic losses have limited their performance. In contrast, superconducting radio frequency (SRF) cavities can achieve quality factors of $\sim \num{1e10}$, perhaps five orders of magnitude better than copper cavities, leading to more sensitive dark matter detectors. In this paper, we first derive that the scan rate of a haloscope experiment is proportional to the loaded quality factor $\ql$, even if the cavity bandwidth is much narrower than the dark matter halo line shape. We then present a proof-of-concept search for dark photon dark matter using a nontunable ultrahigh quality SRF cavity. We exclude dark photon dark matter with kinetic mixing strengths of $\chi > \num{1.5e-16}$ for a dark photon mass of $m_{\Ap} = \SI{5.35}{\mu eV}$, achieving the deepest exclusion to wavelike dark photons by almost an order of magnitude.
\end{abstract}
\maketitle

%\tableofcontents

\section{Introduction}

There is overwhelming evidence that 84.4\% of the matter in the Universe is made out of dark matter (DM)~\cite{Rubin:1982kyu, 10.1093/mnras/249.3.523,1998gravitational_lensing,10.1093/mnras/stw3385, Markevitch_2004, 2020Planck, Zyla:2020zbs}. The $\Lambda$CDM model describes dark matter as feebly interacting, nonrelativistic, and stable on cosmological timescales. Not much else is known about the nature of dark matter, particularly what particles beyond the standard model it is made of.

The dark photon (DP) is a compelling dark matter candidate. It is a spin-1 gauge boson associated with a new Abelian U(1) symmetry and is one of the simplest possible extensions to the Standard Model (SM)~\cite{essig2013dark, PhysRevD.104.092016, PhysRevD.104.095029}. The dark photon, having the same quantum numbers as the SM photon, generically interacts with the SM photon through kinetic mixing~\cite{HOLDOM198665, HOLDOM1986196} described by the Lagrangian
\begin{align}
  \mathcal{L} = -\frac{1}{4}(F_1^{\mu \nu}F_{1\mu \nu} +F_2^{\mu \nu}F_{2\mu \nu} - 2\chi F_1^{\mu \nu}F_{2\mu \nu} - 2 m_{\Ap}^{2} A^{\prime 2}),
  %\label{eqn:dp_mixing_lagrangian}
\end{align}
where $F_1^{\mu \nu}$ is the electromagnetic field tensor, $F_2^{\mu \nu}$ is the dark photon field tensor, $\chi$ is the kinetic mixing strength, $m_{\Ap}$ is the DP mass, and $\Ap$ is the DP gauge field. 
If both $m_{\Ap}$ and $\chi$ are sufficiently small, then it is stable on cosmological timescales~\cite{PhysRevD.78.115012}. The dark photon is then an attractive dark matter candidate. If its mass is less than an eV, the dark photon dark matter (DPDM) is in the wavelike regime, where it is best described as a coherent wave oscillating at the frequency of its rest mass rather than a collection of particles. The dark matter kinetic energy distribution sets the degree of coherence of wavelike dark matter to be of order $v_\mathrm{DM}^2\sim 10^{-6}$~\cite{PhysRevD.42.3572, 10.1046/j.1365-8711.2003.06165.x}.

%The lifetime is about the same as the age of the universe if $m_{A^{\prime}} (\chi^2\alpha)^{1/9} < \SI{1}{keV}$~\cite{PhysRevD.78.115012}, where $\alpha$ is the fine structure constant. This condition is easily met if $m_{\Ap} \approx \SI{e-4}{eV}$ and $\chi < \num{e-12}$.

Several mechanisms could produce a relic of cosmic dark photons. One simple example is the displacement of the DP field through quantum fluctuations during inflation~\cite{PhysRevD.93.103520}. These fluctuations in the DP field serve as the initial displacement for dark photon field oscillations, which commence once the universe's expansion rate $H$ falls below the DP mass, i.e., $H\hbar < m c^2$. Other mechanisms are possible and are described in~\cite{PhysRevD.104.095029, Arias_2012}.

DPDM can be detected through its mixing with the SM photon. If dark photons convert into SM photons inside a microwave cavity with a large quality factor, then a feeble EM signal accumulates inside the cavity, which can be read by ultralow noise electronics. This type of detector is called a haloscope and is often deployed to search for axionic DM~\cite{PhysRevLett.51.1415}. The SM photon frequency $f$ is related to the dark photon energy $E_{\Ap}$ by $hf = E_{\Ap} \approx m_{\Ap}c^2$.

In natural units, the dark photon signal power inside the cavity is~\cite{PhysRevD.104.092016, PhysRevD.106.102002, Kim_2020, PhysRevD.32.2988} 

\begin{align}
  P_{S} = P_0 \betaterm L(\Delta)\label{eqn:dp_power},
\end{align}
where
\begin{align}
  & P_{0} = \begin{cases}
	  \eta \chi^2 m_{\Ap} \rho_{\Ap} \veff \ql, & \text{if $\ql << \qdm$}\\
	  \eta \chi^2 m_{\Ap} \rho_{\Ap} \veff \qdm, & \text{if $\ql >> \qdm$}\\
	  \end{cases}\label{eqn:dp0_power}\\
	& V_{eff} = \frac{\left (\int dV \vb{E}(\vec{x}) \vdot \vb{\Ap}(\vec{x})\right )^2}{\int dV \epsilon_r |\vb{E}(\vec{x})|^2|\vb{\Ap}(\vec{x})|^2},\label{eqn:veff}
\end{align}
$\eta$ is a signal attenuation factor, $\rho_{\Ap}$ is the local density of dark matter, $\veff$ is the effective volume of the cavity, $\ql$ is the cavity's loaded quality factor, $\qdm$ is the dark matter ``quality factor'' related to the dark matter coherence time, and $\beta$ is the cavity coupling coefficient. The Lorentzian term is $L(f, f_0, \ql) = 1/(1+4\Delta^2)$, where $\Delta \equiv \ql (f-f_0)/f_0$ is a detuning factor that depends on the SM photon frequency $f$, cavity resonant frequency $f_0$, and $\ql$. $\veff$ is the overlap between the dark photon field $\vb{\Ap}(\vec{x})$ and the dark photon-induced electric field $\vb{E}({\vec{x}})$. Equations~\ref{eqn:dp0_power}, and \ref{eqn:veff} assume that the cavity size is much smaller than the DP de Broglie wavelength ($\sim \SI{300}{m}$). 

 The dark photon mass is unknown, so haloscopes must be tunable to search through the $\chi$ vs. $m_{\Ap}$ parameter space. Thus, the scan rate is a crucial figure of merit for haloscope experiments. Most haloscope literature has focused on the case where ${\ql << \qdm}$ because copper cavities have been traditionally used. However, superconducting niobium cavities with $\ql \sim \num{1e10}$~\cite{PhysRevApplied.13.034032} are readily available for DPDM haloscope searches, and superconducting cavities resistant to multi-Tesla magnetic fields with $\ql > \qdm$, consisting of dielectrics, Nb$_3$Sn, or HTS tapes, will soon be readily available~\cite{ALESINI2021164641, https://doi.org/10.48550/arxiv.2201.10733, PhysRevApplied.17.054013, PhysRevApplied.17.L061005}. 

This Article first derives that the haloscope scan rate is proportional to $\ql$, even in the case where $\ql >> \qdm$ and the DP signal power saturates (Eq.~\ref{eqn:dp0_power}). This conclusion strongly motivates the pursuit of ultrahigh Q haloscopes. This Article then reports a DPDM search using a nontunable \SI{1.3}{GHz} cavity with $\ql \sim \num{1e10}$. The search demonstrates superior sensitivity enabled by the ultrahigh quality factor and achieves the deepest exclusion to wavelike DPDM to date by almost an order of magnitude.
 %The $(Q_L Q_{DM})/(Q_L + Q_{DM})$ factor is often written as $min(Q_L, Q_{DM})$, but the current form is more accurate for when $Q_{DM} \sim Q_L$~\cite{PhysRevLett.55.1797, Kim_2020}.

\section{The scan rate for an ultrahigh Q haloscope}
The haloscope scan rate is strongly dependent on the signal-to-noise ratio (SNR), where $\snr = (P_S/P_n)\sqrt{b \Delta t}$~\cite{doi:10.1063/1.1770483, Peng:2000hd}. $P_n$ is the noise power, $b$ is the frequency bin width, and $\Delta t$ is the integration time. $P_n$ is the combination of the cavity's blackbody radiation and the receiver's Johnson noise. The noise power can be expressed as $P_n= k_b b T_n$, where $k_b$ is the Boltzmann constant, and $T_n$ is the system noise temperature referenced to the cavity output. 

It is common for a microwave haloscope experiment to implement a circulator at the same temperature as the cavity. For such a system, $T_n$ is constant and independent of the cavity detuning $\Delta$ and cavity coupling $\beta$~\cite{ALKENANY201711}. In other words, the noise temperature is expected to be the same inside and outside the cavity bandwidth.

%\emph{The scan rate for an ultra-high Q haloscope.}---The dark photon mass is unknown, so haloscopes must be tunable to search through the $\chi$ vs. $m_{\Ap}$ parameter space. The scan rate for haloscope experiments is a key figure of merit strongly dependent on the SNR. The SNR for a haloscope signal is $\snr =  P_S/ \sigma_{P_n}$, where $P_n$ is the noise power and $\sigma_{P_n}$ is the RMS fluctuations of the noise power~\cite{https://doi.org/10.48550/arxiv.1807.09369}. $P_n$ is the combination of the cavity's blackbody radiation and the receiver's Johnson noise. The noise power can be expressed as $P_n= k_b b T_n$, where $k_b$ is the Boltzmann constant, and $T_n$ is the system noise temperature referenced to the cavity. If the power fluctuations are Gaussian, then it can be shown that $\snr = (P_S/P_n)\sqrt{b \Delta t}$~\cite{doi:10.1063/1.1770483, Peng:2000hd}, $b$ is the frequency bin width, and $\Delta t$ is the integration time.  
If $\ql >> \qdm$, the cavity width is smaller than the dark matter halo line shape width $\dfdm$. The resulting dark matter signal will follow the Lorentzian cavity response with bandwidth $\Delta f_c = f_0/\ql$. Fortunately, a haloscope is sensitive to a distribution of possible dark photon rest masses corresponding to the cavity resonant frequency $f_c$. In other words, a single cavity tuning step can probe the entire dark matter line shape bandwidth, and the tuning steps need only to be comparable to $\dfdm$. However, this scanning strategy would be sensitive to the virialized portion of dark matter. Any component of dark matter with a sharper dispersion could remain undetected by scanning in steps of $\dfdm$.

The frequency bin width $b$ is typically chosen to be comparable to the dark matter signal bandwidth. Typical haloscope experiments use copper cavities with $\ql << \qdm$, so $b \sim \dfdm = f_0/\qdm$. However, if $\ql >> \qdm$, the signal bandwidth is the same as the cavity bandwidth and $b \sim f_0/\ql$. Thus, the noise power is inversely proportional to the $\ql$, i.e., $P_n \sim k_b \left(f_0/\ql\right) T_n$. The higher $\ql$ is, the lower the noise power.

An estimate of the integration time can be obtained by rearranging the SNR equation ${\Delta t = 1/b \left(\snr \times P_n/P_S \right)^2}$.  The tuning steps are $\Delta f \sim f_0/\qdm$. Putting all this together, the instantaneous scan rate for a dark photon haloscope consisting of an ultrahigh Q microwave cavity, i.e., $\ql >> \qdm$, is

\begin{align}
  \dv{f}{t} = \frac{\Delta f}{\Delta t} \sim \ql \qdm \left (\frac{\eta \chi^2 m_{\Ap} \rho_{\Ap} \veff \beta}{ T_n(\beta+1)\snr}\right )^2. \label{eqn:scan_rate}
\end{align}
Note that Equation~\ref{eqn:scan_rate} does not include the dead time of the experiment associated with tuning the cavity and other operations like rebiasing quantum amplifiers. 

The scan rate equation, Equation~\ref{eqn:scan_rate}, happens to be the same whether $\ql >> \qdm$ or $\ql << \qdm$~\cite{OrpheusPRL}. In both cases, the scan rate is directly proportional to $\ql$~\footnote{Ref.~\cite{Kim_2020} also addresses the scan rate for ultrahigh Q haloscopes, but our treatment differs in a few major ways. For reasons explained in the text, we set the tuning step to $\Delta f \sim f_0/\qdm$ instead of $\Delta f \sim f_0/\ql$ and $b \sim f_0/\ql$ instead of $b \sim f_0/\qdm$. Finally, the derived $T_n$ in Ref.~\cite{Kim_2020} does not appear to apply systems ubiquitous to microwave haloscope experiments that implement circulators between the cavity and first-stage amplifier. For such a system, $T_n$ is independent of $\beta$~\cite{ALKENANY201711}. This independence is recognized in Fig.~5 but is not reflected in their equations.}. 

As a comparison, ADMX uses copper cavities with $\ql \sim \num{8e4}$~\cite{PhysRevLett.127.261803}, whereas niobium SRF cavities can achieve $\ql \sim \num{1e9}- \num{1e11}$ depending on the temperature and cavity treatment~\cite{PhysRevApplied.13.034032}. This suggests that SRF cavities can increase the instantaneous scan rate of haloscope experiments by as much as a factor of $\num{1e5}$.

\section{Dark Photon Dark Matter Search with an SRF Cavity}The Superconducting Quantum Materials and Systems (SQMS) Center, hosted by Fermilab, performs a wide range of multidisciplinary experiments with SRF cavities for quantum computing and quantum sensing. One of these efforts includes a family of SRF haloscope experiments known as SERAPH (SupERconducting Axion and Paraphoton Haloscope). The first phase of SERAPH is a proof-of-principle dark photon search using existing accelerator SRF cavities. Using an SRF cavity with ${\ql \approx \num{4.7e9}}$, a HEMT amplifier, and the standard haloscope analysis is enough to demonstrate superior sensitivity to wavelike dark photons compared to previous searches.

The haloscope consists of a TESLA-shaped single-cell niobium cavity~\cite{Aune:2000gb} with TM$_{010}$ resonant frequency ${f_0 \approx \SI{1.3}{GHz}}$. The cavity is made of fine-grain bulk niobium with a high residual resistivity ratio of $\simeq 300$. The cavity volume is \SI{3875}{mL}, and the effective volume calculated from Equation~\ref{eqn:veff} is $\veff = \SI{669}{mL}$, assuming a randomly polarized DP field. Electromagnetic coupling to the cavities is performed using axial pin couplers at both ends of the beam tubes.

The cavity underwent heat treatments in a custom-designed oven to remove the niobium pentoxide (Nb$_2$O$_5$) and to mitigate the two-level system (TLS) dissipation~\cite{PhysRevLett.119.264801, PhysRevApplied.13.014024}. The central cell \SI{1.3}{GHz} cavity is heat treated at $\sim 450$\degree C in vacuum for one hour. 

The cavity is cooled to $\approx\SI{33}{mK}$ using a BlueFors XLD 1000 dilution refrigerator. A double-layer magnetic shielding around the entire cryostat is used, and magnetometers placed directly on the outside cavity surfaces indicate that the DC ambient magnetic field level is shielded to below \SI{2}{mG}. 

\begin{figure}
  \centering
  \includegraphics[width=0.9\linewidth]{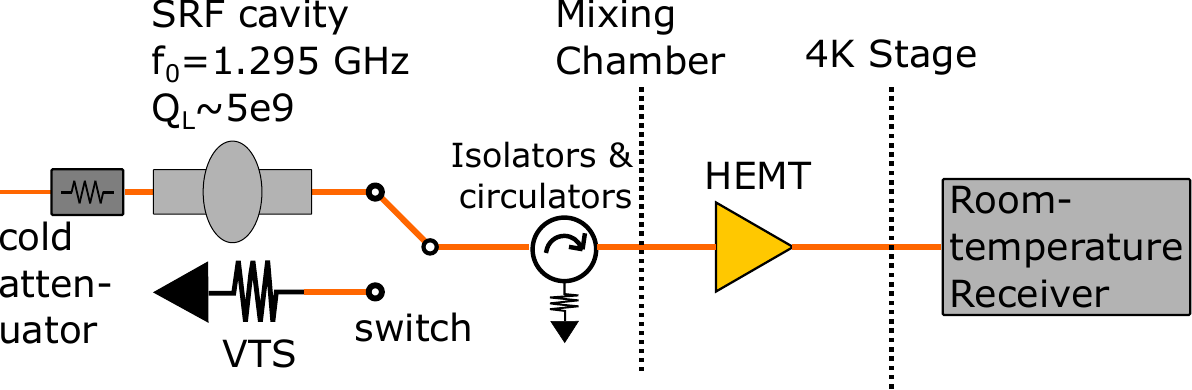}
	\caption{A conceptual diagram of the dark photon dark matter search using an SRF cavity. Power from the cavity is amplified by a cryogenic HEMT amplifier  before reaching the room-temperature receiver. A series of circulators prevent the cavity from being excited by the amplifier's noise. The variable temperature stage (VTS) is used to perform the noise calibration. A technical schematic of the experiment is shown in Fig.~\ref{fig:dr_electronics}.}
  \label{fig:electrtonics}
\end{figure}

A conceptual diagram of the microwave electronics is shown in Fig.~\ref{fig:electrtonics}, and a technical drawing is shown in Fig.~\ref{fig:dr_electronics}. A series of attenuators on the cavity input line attenuates the room-temperature noise. The power from the cavity is first amplified by a cryogenic HEMT amplifier. Four circulators are installed between the HEMT and SRF cavity. The circulators prevent the HEMT amplifiers from injecting noise into the cavity. The signal is further amplified at room temperature and then injected into the appropriate measurement device (spectrum analyzer, network analyzer, or phase noise analyzer).

The system noise temperature referenced to the cavity output, $T_n$, is measured using the Y-factor method~\cite{10.1063/5.0028951,pozar}. For this measurement, the cryogenic switch in Fig.~\ref{fig:electrtonics} is connected to a Variable Temperature Stage (VTS) instead of the SRF cavity. The VTS consists of a \SI{50}{\ohm} load, a resistive heater, and a Cernox thermometer mounted on a copper plate.  The VTS is anchored to the mixing chamber plate via stainless steel standoffs. In this configuration, the system noise power, $P_0$,  depends linearly on the VTS temperature, $\tvts$, via $P = k_b G b (\tvts + \tadd)$, where $G$ is the system gain and $\tadd$ is the added noise from the electronics and includes the insertion loss from the transmission cables and circulators. $\tvts$ is heated from $\SI{.16}{K}$ to $\SI{7.6}{K}$, and the power out of the HEMT is monitored with a spectrum analyzer. From the Y-factor measurement, it is determined that $\tadd = \SI{7.1 \pm 0.4}{K}$. This number is consistent with a HEMT noise temperature of \SI{4.6}{K} and a \SI{2}{dB} insertion loss from the circulators at the mixing chamber temperature. When the switch is connected to the cavity, $T_n = \tcav + \tadd$. During the experiment, $\tcav = \SI{33}{mK}$, leading to $T_n = \SI{7.1 \pm 0.4}{K}$ (Boson statistics need not be considered in the Rayleigh-Jeans limit, $k_b \tcav >> hf$). 

The cavity's resonant frequency is identified using a self-excited loop (SEL)~\cite{Fong2011SELFEO, delayen1978phase}. The thermal noise from the output of the cavity is amplified, phase shifted, and fed back into the input of the cavity. A power splitter feeds the cavity's output power to the spectrum analyzer to monitor the response to the SEL. The peak of the power spectrum corresponds to the cavity resonance. 

The SEL was also used to characterize cavity microphonics, i.e., detuning induced by vibrations. Microphonics causes the cavity's instantaneous frequency to fluctuate over time, which causes the dark matter signal to spread beyond the cavity bandwidth. This frequency modulation causes the dark matter signal power to leak into different sidebands. Two prominent sidebands were identified at \SI{14.3}{Hz} and \SI{57.2}{Hz}. Their combined effect caused the dark matter signal amplitude to be reduced by \SI{0.54}{dB}. The signal attenuation factor from Equation~\ref{eqn:dp0_power} is thus $\eta = 0.88$. The SEL implementation, measurement, and characterization of microphonics are described in more detail in Appendix~\ref{app:microphonics}.

The cavity's loaded quality factor $\ql$ is measured using a decay measurement~\cite[p.155]{padamsee}, where the cavity is first energized and the time constant $\tau_L$ in which the energy decays is extracted. The decay measurements consistently demonstrated $Q_L = 2\pi f_0 \tau_L = \num{4.7e9}$. The antenna external quality factors $Q_{e1}$ and $Q_{e2}$ are measured beforehand in a separate liquid helium bath test stand following the procedure outlined in Reference~\cite{doi:10.1063/1.4903868}. The external quality factors were determined to be $Q_{e1}=\num{8.4e9}$ and  $Q_{e2}=\num{2.3e12}$ with 10\% uncertainty. This results in an unloaded Q, $Q_0 = \num{1.1e10}$. The cavity coupling coefficient of the output port is determined to be $\beta = \left(\ql/Q_e\right)/\left(1-\ql/Q_e \right) = \num{1.3\pm0.3}$.

For this proof-of-principle measurement, there is no tuning mechanism. A single power spectrum is measured. In the absence of a discovery, an exclusion on the kinetic mixing strength $\chi$ is determined from the measured power spectrum, the system noise temperature, and cavity properties. The relevant properties for determining the dark photon signal power and system noise temperature are shown in Table~\ref{tab:operating_parameters}.

\begin{table}[htp]
\begin{tabular}{c|c|c|c|c|c|c|c}
	$\eta$ & $\beta$ & $\veff$ & $m_{\Ap}$ & $\Delta t$ & $\ql$ & $T_n$ & b \\ \hline 
	\num{0.88} & \num{1.3 \pm 0.3} & \SI{669}{mL} & \SI{5.4}{\mu eV} & \SI{1000}{s}& \num{4.7e9} & \SI{7.1 \pm 0.4}{K} & \SI{0.1}{Hz}
\end{tabular}
\caption{Operating parameters for the dark photon dark matter search with the SQMS SRF cavity.}
\label{tab:operating_parameters}
\end{table}

%The signal attenuation factor $\eta$ is determined from cascaded insertion loss of the three isolators and low-pass filter in between the cavity and cryogenic amplifier (Fig.~\ref{fig:electrtonics}). From the manufacturer datasheets, the maximum combined loss is \SI{2.5}{dB}. This leads to a signal attenuation factor of $\eta=0.56$. Since this is derived from cascading the maximum loss of four devices, this value is treated as a conservative lower bound.

The detector sensitivity is estimated from the operating parameters shown in Table~\ref{tab:operating_parameters}. The detector sensitivity can be estimated by rearranging the SNR equation to solve for the kinetic mixing parameter $\chi$:

\begin{align}
  \chi &= \sqrt{\frac{\beta+1}{\beta}\frac{\snr \times  T_n}{\eta m_{\Ap}\rho_{\Ap}V_{eff}\qdm}}\left    ( \frac{\Delta f_c}{\Delta t} \right )^{1/4}\label{eqn:dp_power_estimate}.
\end{align}

For this estimate, the bandwidth $b$ is set to the cavity bandwidth, $\rho_{\Ap} = \SI{0.45}{GeV/cm^3}$, and $\qdm \approx \num{1e6}$. For this sensitivity estimate, setting $\snr = 2$ approximates a 90\% exclusion limit. The parameters in Table~\ref{tab:operating_parameters} are converted to natural units, and the projected detector sensitivity $\chi_{\text{proj}}$ is estimated to be $\chi_{\text{proj}}= \num{1.9e-16}$.

\begin{figure}
  \centering
  \includegraphics[width=\linewidth]{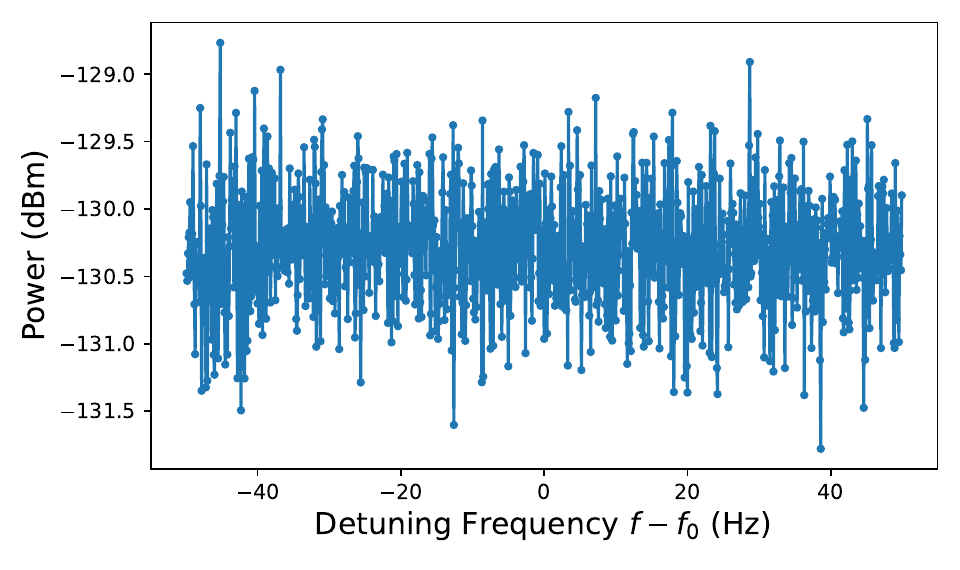}
	\caption{Raw spectrum as measured by the spectrum analyzer. The spectrum bin size is \SI{100}{mHz}, and 100 averages were taken for a total integration time of \SI{1000}{s}. The cavity frequency is $f_0 = \SI{1.294605478}{GHz}$ and the cavity bandwidth is \SI{277}{mHz}. Assuming the spectrum is just noise, the raw spectrum is the noise power of the system.}
  \label{fig:power}
\end{figure}

For the dark photon search, power from the cavity is measured using the Rohde \& Schwarz FSW-26 Signal and Spectrum Analyzer. The cavity $\ql = \num{4.7e9}$, so the frequency resolution needs to be $b \sim \SI{100}{mHz}$. This sub-Hertz resolution is achieved using the spectrum analyzer's I/Q-analyzer mode. These measurements use a \SI{1}{kHz} sample rate and a \SI{10}{s} sweep time. The power spectrum consists of 100 averages, resulting in a total integration time of \si{1000}{s}. The spectrum's center frequency is set to the cavity's resonant frequency of \SI{1.294605478}{GHz}. The frequency bin size is \SI{100}{mHz}. The cavity bandwidth is \si{277}{mHz}, so the cavity spans about \num{2.77} bins. After the data is recorded, the spectrum is truncated further so the span is about \SI{100}{Hz} centered around the cavity resonance. Most of the power spectrum is outside of the cavity bandwidth, allowing for a convenient verification that the noise is Gaussian. On this frequency scale, the power fluctuations are Gaussian without the need for the application of a low pass filter (the Savitzky-Golay filter is typical of haloscope analysis~\cite{PhysRevD.96.123008, PhysRevD.103.032002}). The resulting power spectrum is shown in Fig.~\ref{fig:power}.

Once the power spectrum is measured, the standard haloscope analysis is applied to either find a spectrally narrow power excess consistent with a dark photon signal or to exclude parameter space. The procedure for deriving the exclusion limits follows the procedure developed by ADMX and HAYSTAC~\cite{PhysRevD.64.092003, PhysRevD.96.123008, PhysRevD.103.032002}, and is adapted for dark photon searches~\cite{PhysRevD.104.095029, PhysRevD.104.092016, PhysRevD.106.102002}. 

There are a few important deviations from the standard haloscope analysis for this search. First, only one spectrum was measured, so combining many spectra at different RF frequencies is unnecessary. Second, the frequency range of interest is narrow enough such that the measured power is unaffected by the frequency-dependent gain variation of the electronics. So, the Savitzky-Golay filter is not needed to remove this gain variation. This is advantageous because the Savitzky-Golay filter is known to attenuate the dark matter signal by as much as 10\%-20\%~\cite{PhysRevD.96.123008, PhysRevD.103.032002}. 

Third, most of the data points in Fig.~\ref{fig:power} used to verify the noise's Gaussianity and determine the statistical parameters are well outside the cavity bandwidth. Fortunately, in the absence of a dark matter signal, the statistical distribution of the power fluctuations is the same inside and outside of the cavity bandwidth. 

Fourth, past haloscope experiments with $\ql << \qdm$ typically convolved the spectra with the dark matter halo line shape to account for the signal being spread across multiple bins. For this search, $\ql >> \qdm$, so the signal will be Lorentzian from the cavity response. So the spectrum is convolved with the cavity line shape $L(f, f_0, \ql) = 1/(1+4\Delta^2)$. 

Fifth, a single measurement is sensitive to a range of dark photon masses. Thus, the excluded power on resonance is convolved with the dark matter halo line shape. This convolution was also performed in other dark photon searches with $\ql > \qdm$~\cite{PhysRevLett.126.141302}. When performing this convolution, it should be noted that the photon frequency (which corresponds to the cavity frequency) is fixed, and it is the dark photon mass that varies.

\begin{figure}
  \centering
  \subfloat{\includegraphics[width=\linewidth]{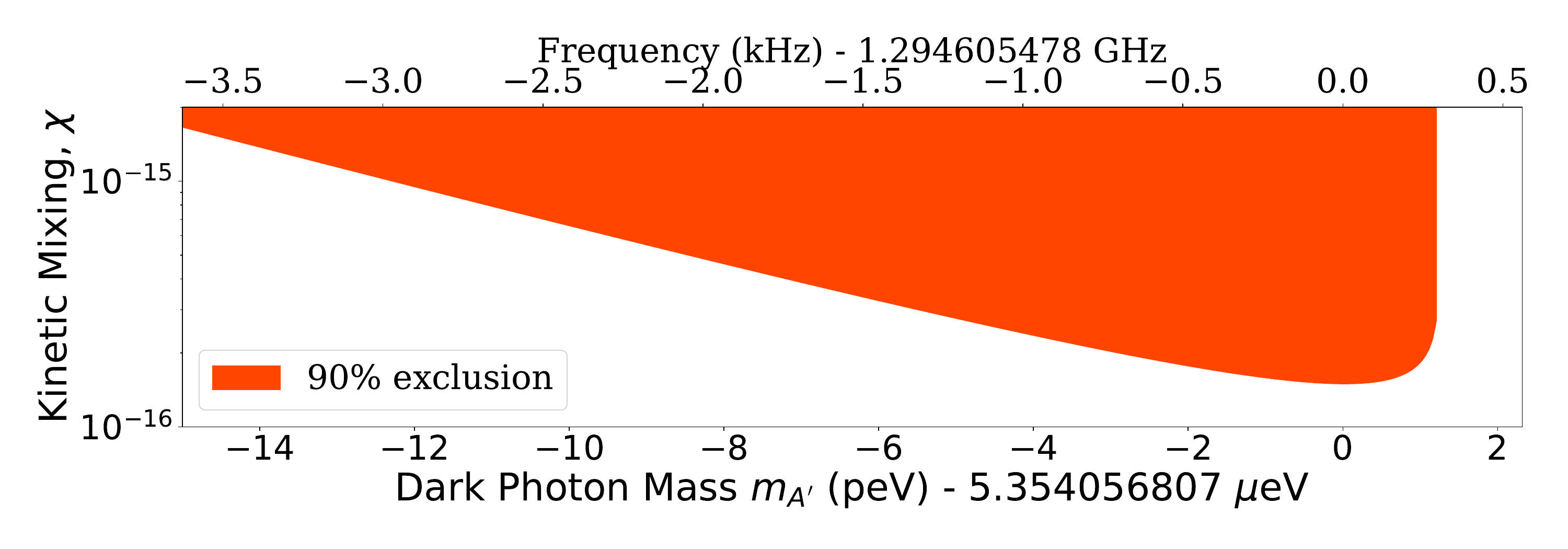}}\\
  \subfloat{\includegraphics[width=\linewidth]{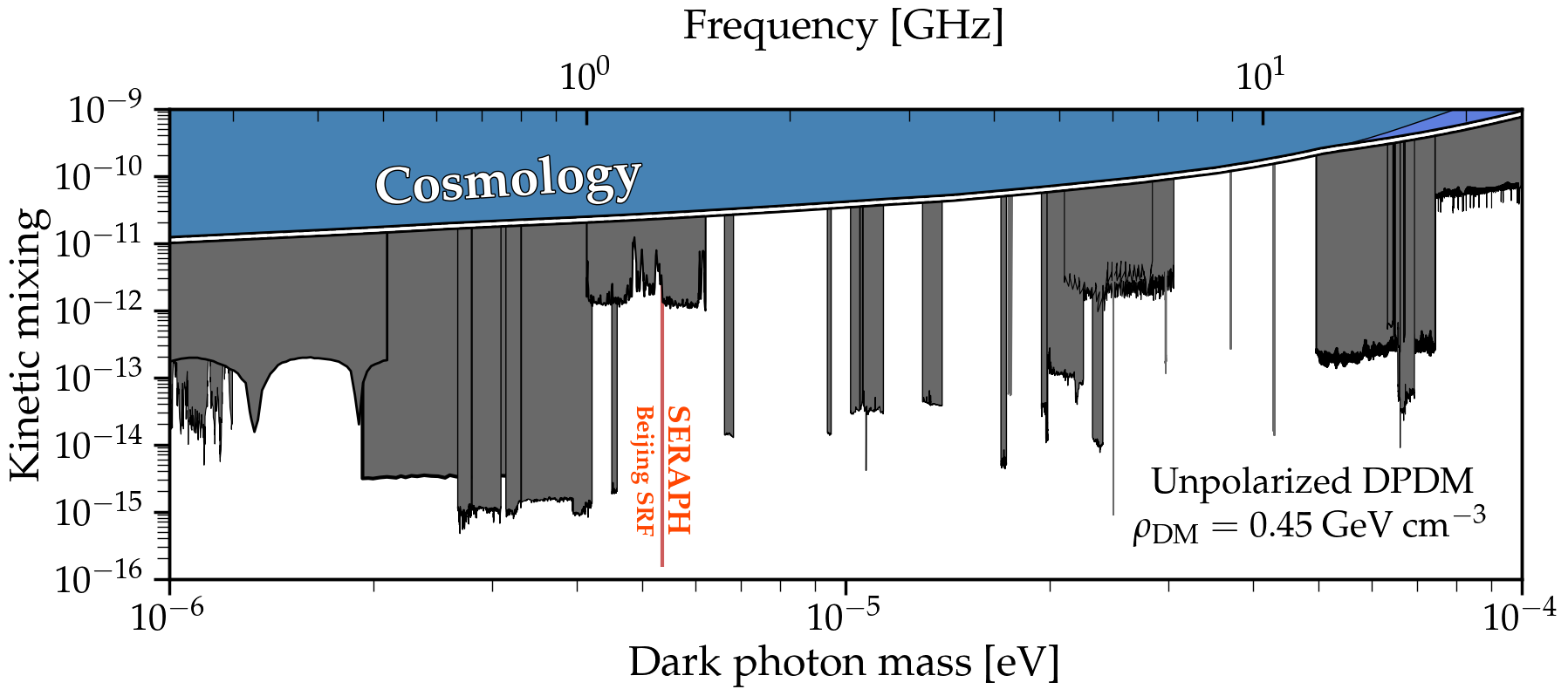}}
  \caption{Top: A 90\% exclusion on the kinetic mixing strength parameter space. Bottom: SQMS limits in the context of other microwave cavity haloscopes. Figure adapted from~\cite{ciaran_o_hare_2020_3932430}.}
  \label{fig:dp_limits}
\end{figure}

No spectrally narrow power excess with an $\snr > 5$ is found in the measured power spectrum. The excluded parameter space using a 90\% confidence limit is shown in Fig.~\ref{fig:dp_limits}. The derived limit assumes dark photon dark matter is randomly polarized, and the dark photon energy distribution follows the standard halo model. The excluded kinetic mixing strength is $\chi_{90\%} = \num{1.5e-16}$ for a dark photon mass of $m_{\Ap} = \SI{5.354}{\mu eV}$. This $\chi_{90\%}$ is consistent with the expected sensitivity estimated from Equation~\ref{eqn:dp_power_estimate}. It is also the deepest exclusion to wavelike dark photon dark matter by almost an order of magnitude.

%Conceptually, the shape of $\chi_{90\%}$ makes sense. Lower mass dark photons have a slight chance of being detected if they have a large enough kinetic energy to match the cavity frequency ($f = E_{\Ap}$). However, dark photons with rest masses higher than the photon energy will not be detected.

\section{Outlook and the Potential of SRF Cavities for Axion Dark Matter Searches}The exclusion in Fig.~\ref{fig:dp_limits} is an impressive demonstration of how SRF cavities can benefit dark matter searches. To extend the range of frequencies in a broader dark matter search, SQMS is currently developing experiments using widely-tunable SRF cavities. 

In addition to dark photons, there is a growing interest in dark matter axions. Axions are particularly well motivated because they solve the strong CP problem~\cite{PhysRevLett.38.1440}. Axion haloscope searches require multi-Tesla magnetic fields to be sensitive enough to the QCD axion. The scan rate for axion haloscope searches is still directly proportional to $\ql$. Unfortunately, the performance of superconductors degrades under an external magnetic field. Achieving high quality factors is thus a very active area of research~\cite{ALESINI2021164641, https://doi.org/10.48550/arxiv.2201.10733, PhysRevApplied.17.054013, PhysRevApplied.17.L061005}, and it seems likely that axion haloscopes with $\ql > \num{1e7}$ are achievable in the near future. 

This experiment also demonstrates that axion haloscopes with $\ql \sim  \num{1e10}$ are worth striving for. Applying a hypothetical \SI{8}{T} magnetic field to the dark photon data in Fig.~\ref{fig:power} would have led to an exclusion on the axion-photon coupling constant at $\gagg \sim \num{3e-16}$, well below DFSZ coupling ($\gagg = \num{8e-16}$). 

Despite decades of searching for the axion, only a small fraction of the QCD axion parameter space has been explored. Perhaps a combination of ultrahigh Q cavities, subSQL metrology~\cite{Backes2021, PhysRevLett.126.141302}, multiwavelength detector designs~\cite{PhysRevLett.129.201301, Brun2019, PhysRevLett.118.091801, PhysRevD.98.035006, PhysRevLett.128.231802, PhysRevApplied.9.014028, PhysRevApplied.14.044051, PhysRevLett.123.141802}, and innovations in multi-Tesla continuous magnets will enable experiments to probe most of the post-inflation QCD axion parameter space within the next few decades.

%\begin{figure}
%  \centering
%  \includegraphics[width=\linewidth]{AxionPhoton_RadioFreqCloseup_noNS}
%  \caption{Top: A 90\% exclusion on the mixing angle parameter space. Bottom: Orpheus limits in the context of other microwave cavity haloscopes. Figure adapted from~\cite{ciaran_o_hare_2020_3932430}.}
%  \label{fig:axion_limits}
%\end{figure}

\emph{Acknowledgements}---The authors thank Asher Berlin, Yonatan Kahn, Akash Dixit, and Benjamin Brubaker for fruitful discussions regarding scanning strategies and data analysis with ultrahigh Q cavities. The authors thank Andrew Penhollow and Theodore C. Ill for the cavity assembly. This material is based upon work supported by the U.S. Department of Energy, Office of Science, National Quantum Information Science Research Centers, Superconducting Quantum Materials and Systems Center (SQMS) under contract number DE-AC02-07CH11359%\FloatBarrier

\FloatBarrier
\appendix
\section{Detailed Schematic}
\begin{figure}
  \centering
  \includegraphics[width=\linewidth]{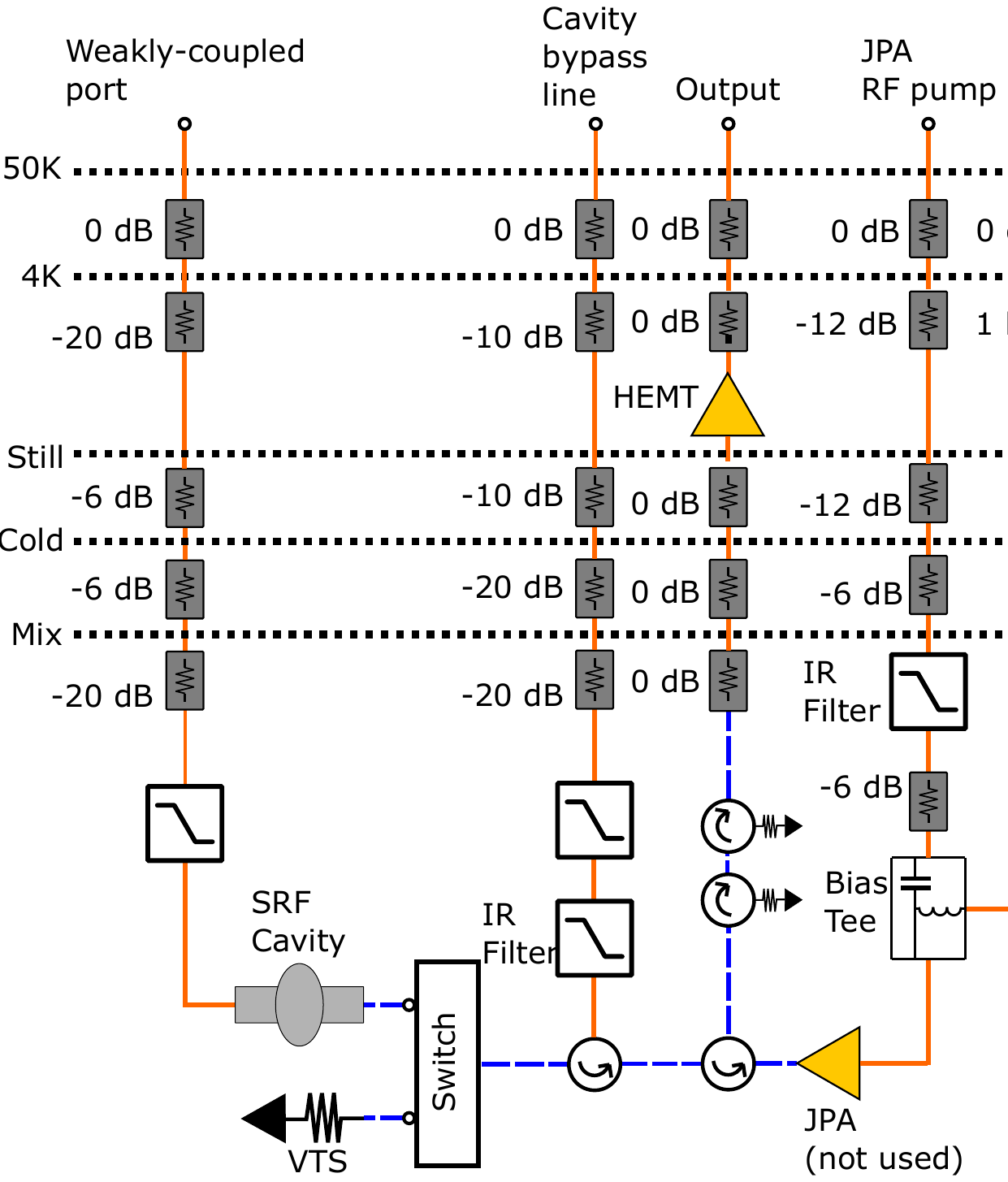}
	\caption{A more detailed schematic of the wiring and microwave electronics in the dilution refrigerator. The blue dashed lines represent superconducting NbTi cables.}
  \label{fig:dr_electronics}
\end{figure}

The conceptual diagram in Fig.~\ref{fig:electrtonics} is simplified to increase the readability of the main text. The complete schematic of the electronics in the dilution refrigerator is shown in Fig.~\ref{fig:dr_electronics}. The cryogenic switch is a Radiall R583423141. The cryogenic HEMT amplifier is from Low Noise Factory (LNF-LNC0.3\_14A~\cite{lnf_0p314}). At \SI{1.3}{GHz}, the manufacturer has measured the amplifier noise temperature at \SI{4.9 \pm 0.5}{K} and the gain at \SI{36}{dB}. The four circulators are QuinStar QCY-G0110151AS. Each circulator is expected to have a \SI{0.5}{dB} insertion loss and 18~dB isolation. The bias tee is a Marki DPXN-M50. A room-temperature low-noise amplifier (Fairview Microwave FMAM1028, not shown) is used to amplify the signal further before reaching the spectrum analyzer.  

A Josephson Parametric Amplifier (JPA) was installed but not used for the dark matter search. The JPA was fabricated by the National Institute of Standards and Technology (NIST). The JPA gain is maximized at \SI{1.9}{GHz} and had a tuning range between \SI{1.2}{GHz} and \SI{2.5}{GHz}, but otherwise this JPA was never fully characterized. During this cooldown, the JPA was measured to have an overall gain of \SI{3}{dB} at \SI{1.3}{GHz}. It was decided that this additional gain was not worth the added complexity and systematic uncertainties in the noise calibration. 

\section{Decay measurement}
\begin{figure}
  \centering
  \includegraphics[width=\linewidth]{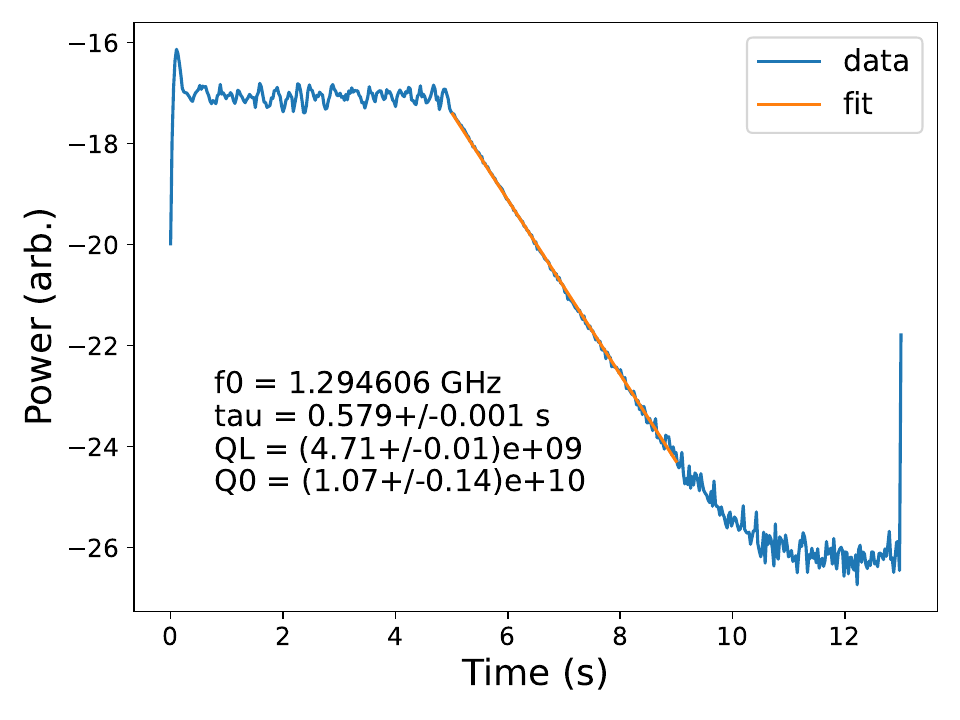}
	\caption{The decay measurement of the \SI{1.3}{GHz} cavity used to derive the loaded quality factor.}
  \label{fig:decay_measurement}
\end{figure}

The decay measurement is implemented with a Rohde \& Schwarz ZNA in the pulse mode. At the resonant frequency, the network analyzer injects a \SI{15}{dBm} signal with a bandwidth of \SI{200}{Hz} into the input transmission line. Port 2 of the VNA measures the absolute power from the cavity output line. The network analyzer source is then turned off, and the output power is observed to decay over several seconds until it reaches an equilibrium. The measured power is fitted using $P_t = A\exp(-t/\tau_L)$, and $\ql = 2\pi f_0 \tau_L$. The result is shown in Fig~\ref{fig:decay_measurement}.

\section{Self-excited Loop and Microphonics}\label{app:microphonics}
\begin{figure}
  \centering
  \includegraphics[width=\linewidth]{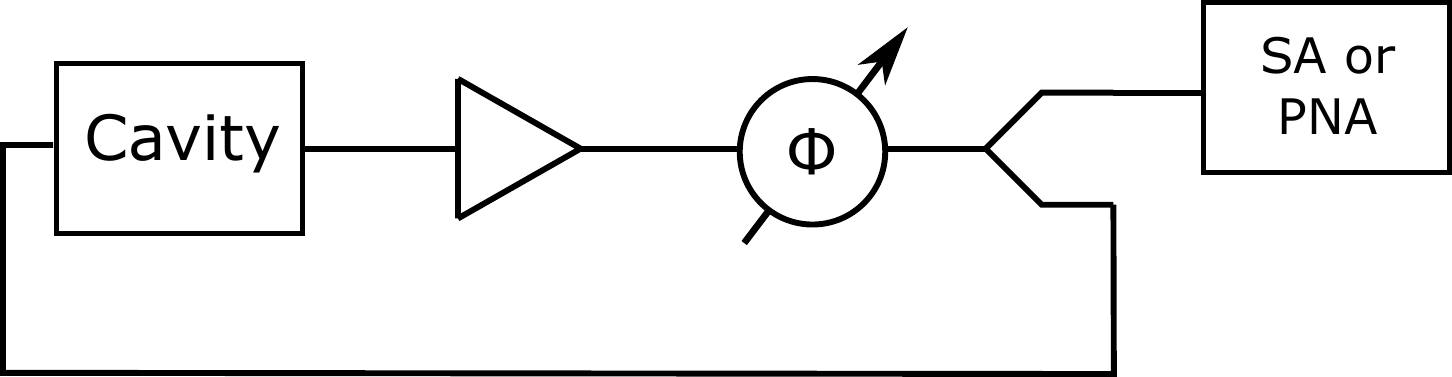}
  \caption{A schematic for the self-excitation loop used to find the cavity's resonant frequency and characterize microphonics.}
  \label{fig:sel}
\end{figure}

A conceptual diagram of the self-excited loop (SEL) is shown in Fig.~\ref{fig:sel}. The phase shifting is performed with an ATM PNR P2214. A power splitter feeds the cavity's output power to the spectrum analyzer to monitor the response to the SEL. In the real implementation, three amplifiers were used: one cryogenic HEMT, the Fairview Microwave FMAM1028 LNA, and a Minicircuits 15542 Model No. ZHL-42. A directional coupler was used along with the power splitter so that the spectrum analyzer and phase noise analyzer could monitor the SEL simultaneously. A bandpass filter (Lorch 6BC-1300/75-S) was used to reduce broadband noise, and an isolator was used between the Fairview LNA and Minicircuits amplifier to mitigate unwanted self-oscillations.

\begin{figure}
  \centering
  \includegraphics[width=\linewidth]{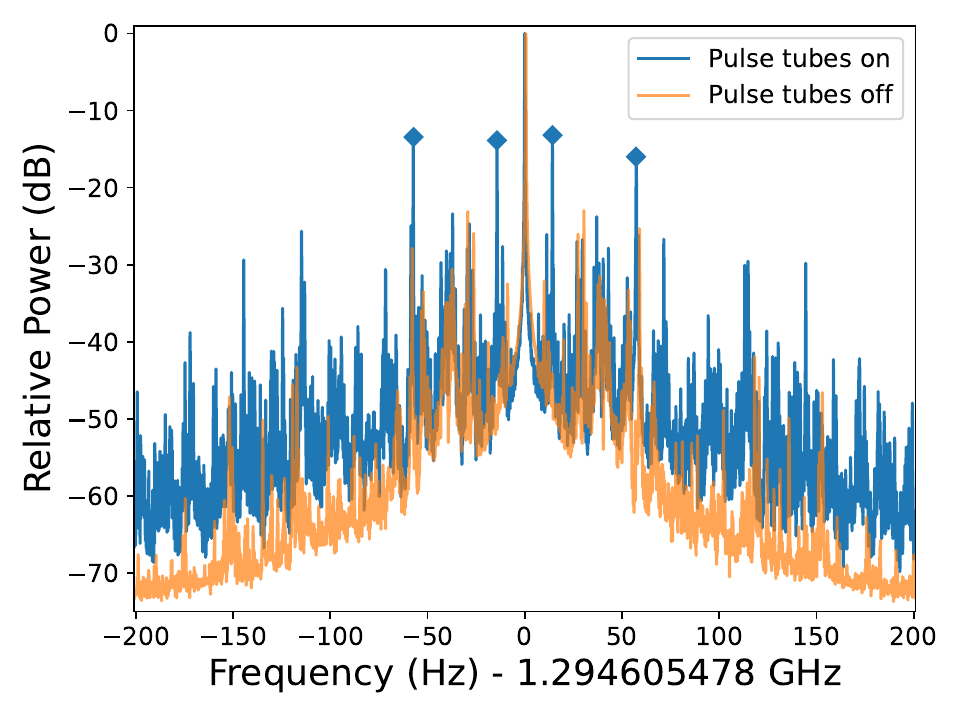}
	\caption{The power spectrum coming out of cavity from self-excitation loop. Microphonics, mostly from the dilution refrigerator pulse tubes, introduces a modulating effect. The sideband amplitude was greatly diminished by turning off the pulse tubes. The 14.3~Hz and 57.2~Hz sidebands discussed extensively in the text are marked with diamonds.}
  \label{fig:microphonics_sa}
\end{figure}

The resulting power is shown in Fig.~\ref{fig:microphonics_sa}. The central peak corresponds to the resonant frequency. There are prominent sidebands spaced \SI{14.3}{Hz} and \SI{57.2}{Hz} apart from the central peak and a forest of smaller peaks; both sidebands are roughly \SI{15}{dB} smaller than the carrier frequency. The power within the carrier peak and sidebands are confined within the cavity bandwidth. These sidebands are caused by the modulation of the resonant frequency due to microphonics. These vibrations originate primarily from the dilution refrigerator's pulse tubes. Fig.~\ref{fig:microphonics_sa} demonstrates that these sidebands are mitigated when the pulse tubes are turned off.

\begin{figure}
  \centering
  \includegraphics[width=\linewidth]{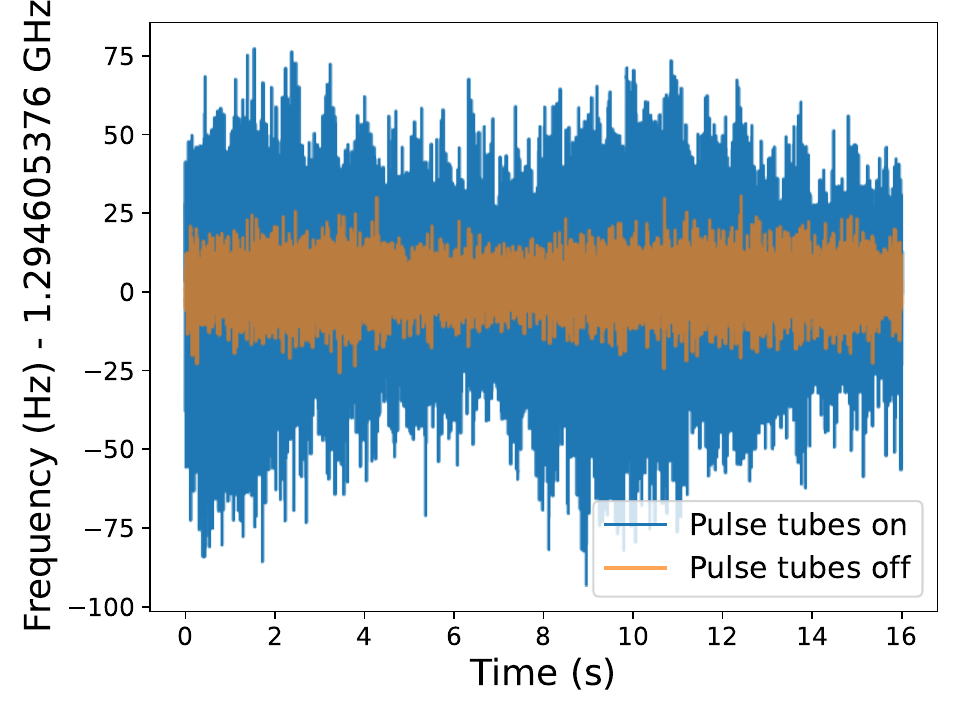}
  \caption{The instantaneous cavity frequency as a function of time, as measured by a phase noise analyzer.}
  \label{fig:microphonics_pna}
\end{figure}

Microphonics affects the dark matter search sensitivity by spreading the potential dark matter signal from the central cavity frequency into sidebands. The frequency modulation framework is useful for understanding the microphonics and the effects on the potential dark matter signal. The Rohde \& Schwarz FSWP Phase Noise Analyzer is used to measure the cavity frequency as a function of time (Fig.~\ref{fig:microphonics_pna}). The root-mean-square of the frequency deviation is \SI{24.7}{Hz}. 

\begin{figure}
  \centering
  \includegraphics[width=\linewidth]{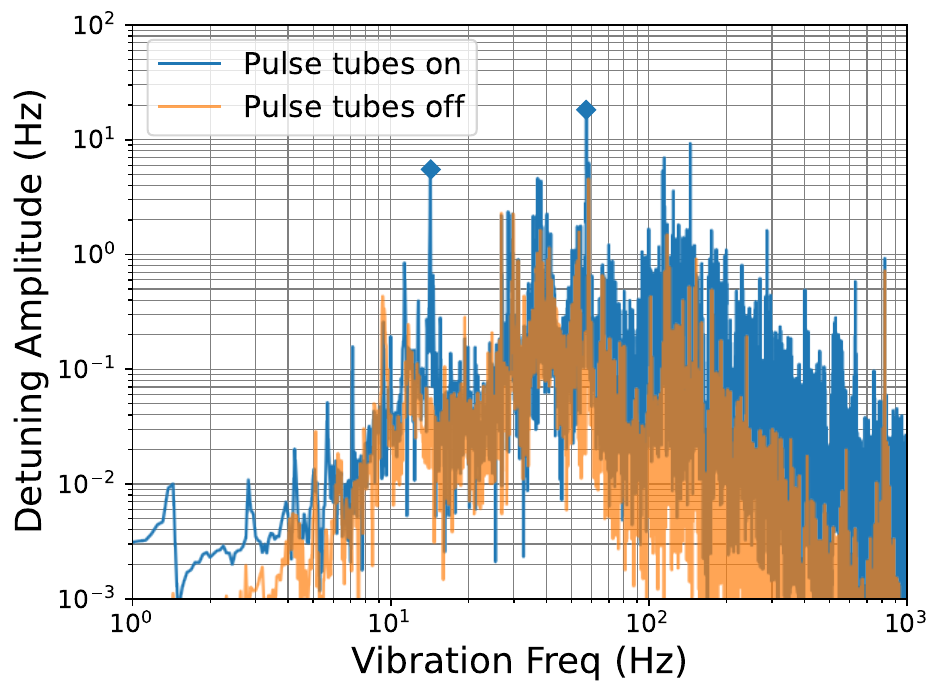}
	\caption{The Fourier Transform amplitude of the microphonics measurement from Fig.~\ref{fig:microphonics_pna}, revealing the cavity vibration frequencies and their corresponding detuning amplitudes. The 14.3~Hz and 57.2~Hz sidebands discussed extensively in the text are marked with diamonds.}
  \label{fig:microphonics_pna_fft}
\end{figure}

As demonstrated in Fig.~\ref{fig:microphonics_pna_fft}, taking the amplitude of the Fourier Transform of the cavity frequency reveals the modulation frequencies $f_m = (\SI{14.3}{Hz}, \SI{57.2}{Hz})$. The corresponding cavity detuning amplitude, i.e., frequency deviation, is $f_{\Delta} = (\SI{5.5}{Hz}, \SI{18.2}{Hz})$. The modulation indices  are $h = f_{\Delta}/f_m = (0.4, 0.3)$, which correspond to sideband amplitudes of (\SI{-14.5}{dBc}, \SI{-16.1}{dBc}) and is consistent with what is observed in Fig.~\ref{fig:microphonics_sa}. There are other notable modulation frequencies in Fig.~\ref{fig:microphonics_pna_fft}, but they have smaller modulation indices, so their corresponding sideband amplitudes are subdominant to the \SI{14.3}{Hz} and \SI{57.2}{Hz} sidebands. One can see from Fig.~\ref{fig:microphonics_sa} that the \SI{37}{Hz} modulation frequency corresponding to sideband amplitude \SI{10}{dB} smaller than the \SI{14.3}{Hz} and \SI{57.2}{Hz} sidebands. Other sidebands are even more subdominant.

The modulation frequencies (\SI{14.3}{Hz}, \SI{57.2}{Hz}) with detuning amplitudes (\SI{5.5}{Hz}, \SI{18.2}{Hz}) correspond to a reduction in the carrier amplitude by (\SI{0.22}{dBc}, \SI{0.32}{dBc}). This leads to a total reduction of the DM signal by \SI{0.54}{dB}, corresponding to a dark matter signal attenuation factor of $\eta = 0.88$. This reduction in the carrier amplitude is confirmed numerically by taking the FFT of 
\begin{align*}
	y(t) = \cos(2\pi f_c t + \frac{f_{\Delta 1}}{f_{m1}} \sin(2\pi f_{m1}t) + \frac{f_{\Delta 2}}{f_{m2}} \sin(2\pi f_{m2}t)), 
\end{align*}
where $f_c = \SI{100}{Hz}$, $f_{\Delta 1} = \SI{5.5}{Hz}$, $f_{m 1} = \SI{14.3}{Hz}$, $f_{\Delta 2} = \SI{18.2}{Hz}$, and $f_{m 2} = \SI{57.2}{Hz}$~\cite{Faruque2017}. This numerically-calculated FFT also confirms the amplitude of the sidebands.

There was an initial attempt to mitigate the effects of microphonics on the dark matter signal by turning off the pulse tubes during the dark matter search. The temperatures of both the cavity temperature and dilution refrigerator mixing chamber are stable for more than \SI{20}{min.} immediately after the pulse tubes are off. Unfortunately, the HEMT amplifier's physical temperature rises from \SI{2}{K} to \SI{9}{K} during this period, making the noise calibration questionable. Thus, the dark matter search incorporating microphonics is used for this publication.

One might make a more sophisticated analysis that searches for the dark matter signal lost in the sidebands to improve sensitivity, but this has yet to be incorporated into this analysis.

\section{Cavity stability}
\begin{figure}
  \centering
  \includegraphics[width=\linewidth]{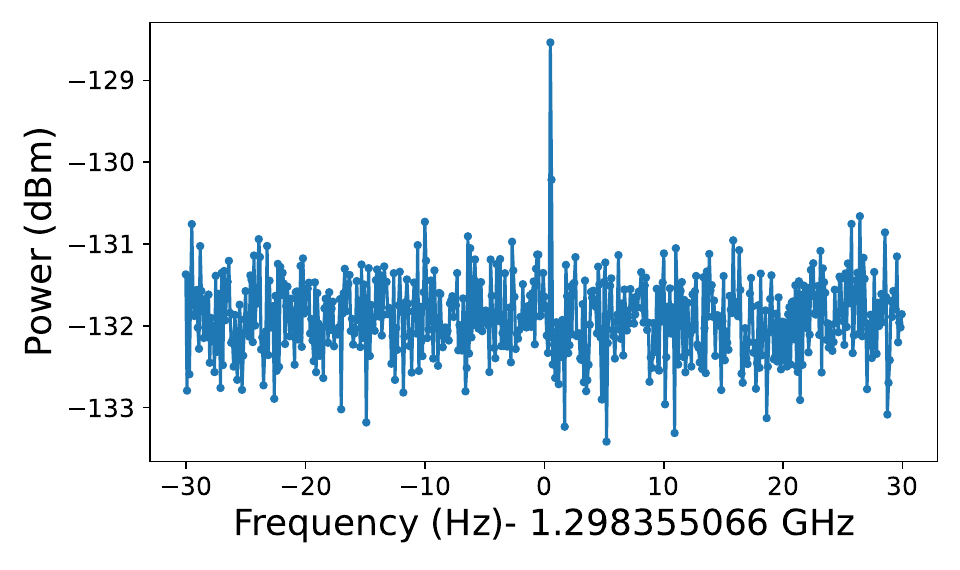}
	\caption{The power spectrum of a \SI{1.3}{GHz} cavity being excited by the noise from HEMT amplifiers. The bin size is \SI{100}{mHz} and the cavity bandwidth for this measurement \SI{433}{mHz}.}
  \label{fig:hemt_excitation}
\end{figure}

A previous experiment by SQMS (a material property study of an SRF cavity not intended for dark matter searches) demonstrates that the cavity is stable for \SI{1000}{s}. The previous setup was similar to Fig.~\ref{fig:dr_electronics}, except that the cavity was connected directly to a HEMT (with three \SI{0}{dB} attenuators in between for thermalization). There was no switch, cavity bypass line, JPA line, or series of circulators. That means that the noise from the HEMT was being directly injected into the cavity, exciting the cavity. The resulting power spectrum is shown in Fig.~\ref{fig:hemt_excitation}. There is a power excess that corresponds to the cavity frequency. A different cavity is used here with $\ql = \num{3.07e9}$, corresponding to a cavity bandwidth of \SI{433}{mHz}. The power spectrum was taken with a bin size of \SI{100}{mHz}. One hundred averages were taken, resulting in a total integration time of \SI{1000}{s}. The bandwidth of the power excess corresponds to the bandwidth of the cavity. This measurement thus shows that the cavity central frequency is stable on this timescale.

A future study to determine the stability of the cavity frequency would be to measure the Allan standard deviation of the cavity frequency.

\section{Noise calibration}
\begin{figure}
  \centering
  \includegraphics[width=0.6\linewidth]{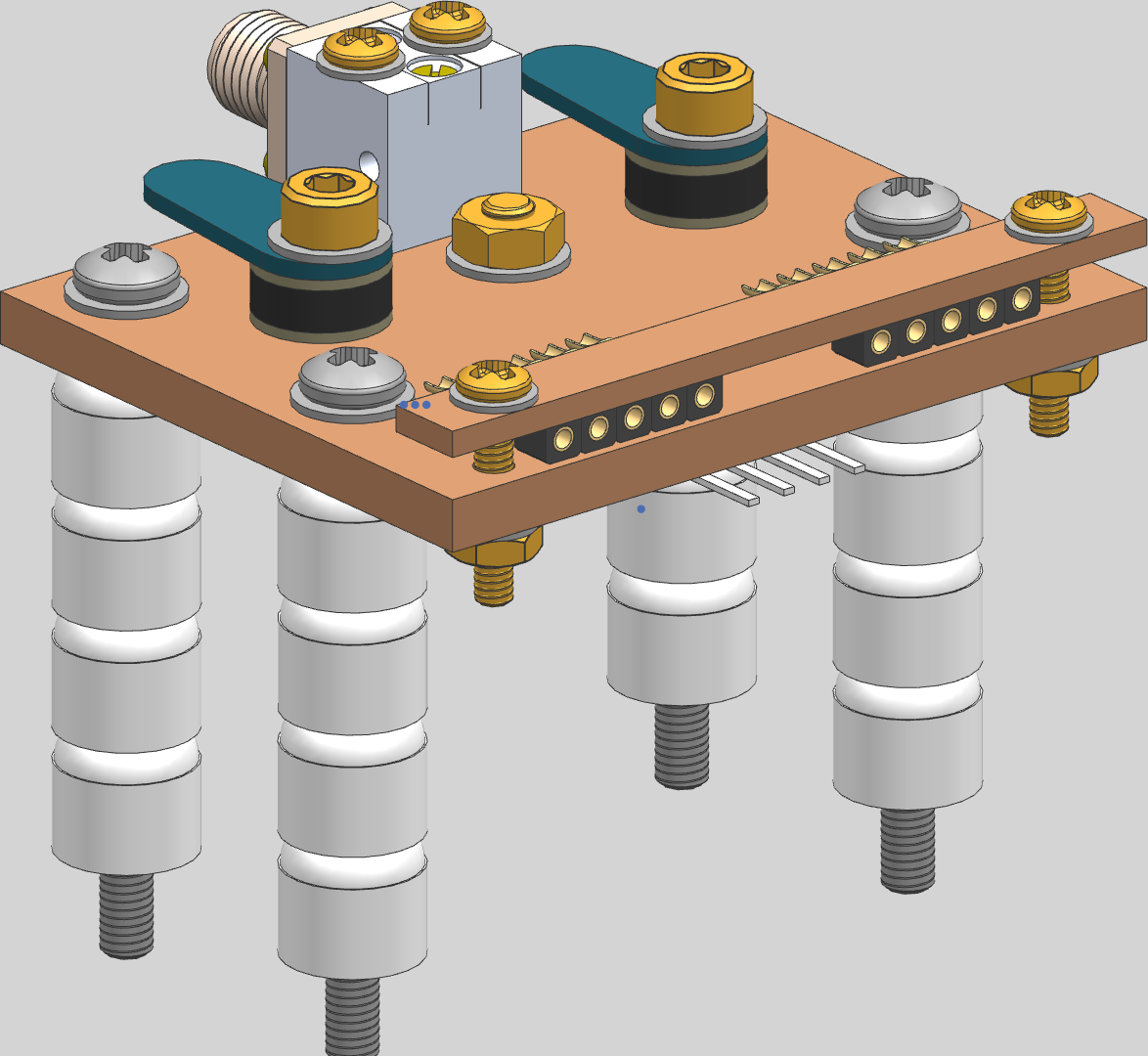}
	\caption{The variable temperature stage is used to perform the noise calibration. It consists of a \SI{50}{\ohm} matched load and a resistive heater anchored to a copper plate.}
  \label{fig:vts}
\end{figure}

The Variable Temperature Stage (VTS) is shown in Fig.~\ref{fig:vts}. It consists of a \SI{50}{\ohm} matched load anchored to a $\SI{50}{mm}\times \SI{41}{mm} \times \SI{3.2}{mm}$ copper plate. A power \si{100}{\ohm} VPR221 power resister from Vishay Precision Group heats the copper plate and is anchored to the other side of the copper plate. Two Cernox temperature sensors (CX-1010-CU-HT-0.1L) are used to measure the temperature of the copper plates. The VTS is attached to the copper brackets surrounding the \SI{1.3}{GHz} cavity via stainless steel standoffs. Alumina fish beads surround the stainless steel standoffs. Alumina has a higher thermal conduction than stainless steel above \SI{3}{K}, allowing for quicker cooldowns.

\begin{figure}
  \centering
  \includegraphics[width=\linewidth]{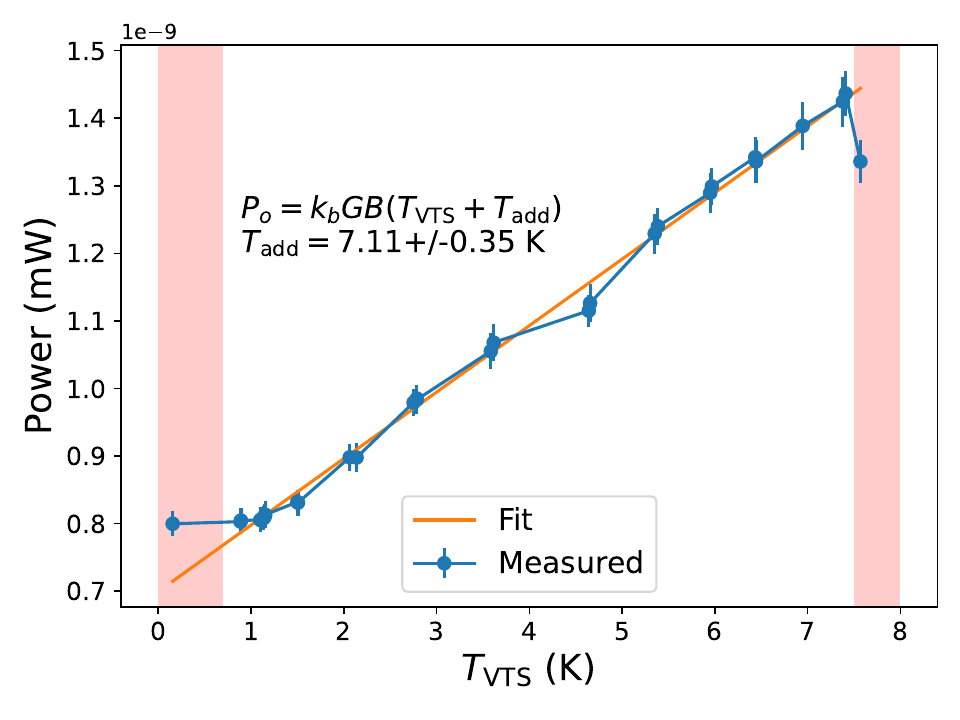}
  \caption{The noise calibration is performed using the Y-factor method, where the noise power of the electronics chain is measured as a function of a varying noise source. The shaded red regions are excluded from the noise calibration fit.}
  \label{fig:yfactor}
\end{figure}

The noise calibration is performed by varying the VTS temperature $\tvts$ from base temperature to \SI{7.6}{K} and monitoring the power out of the HEMT with a spectrum analyzer. The spectrum analyzer is centered at the cavity resonant frequency and is set to a \SI{100}{kHz} span with \num{1000} sweep points and a \SI{1}{kHz} resolution bandwidth (RBW). The noise power does not vary appreciably over this frequency range, so the noise power at a particular $\tvts$ is taken to be the mean of the sweep points, and the uncertainty is taken to be the standard deviation. The measured noise power is plotted in Fig.~\ref{fig:yfactor}.

The linear fit to extract $\tadd$ is also shown in~\ref{fig:yfactor}. The datapoints $\tvts < \SI{0.2}{K}$ and $\tvts > \SI{7.5}{K}$ are excluded from the fit. At $\tvts > \SI{7.5}{K}$, the NbTI cables are losing their superconductivity, which increases attenuation through the cables and, consequently, the measured noise power. It is unclear why there the noise power plateaus $\tvts < \SI{0.2}{K}$. Perhaps the excitation current for the resistance measurement leads to self-heating. However, the data points where $\tvts$ approaches $\tadd$ are most relevant and reliable for the noise calibration. 

\FloatBarrier
\bibliography{sqms_ultrahighq_haloscope}% Produces the bibliography via BibTeX.

\begin{thebibliography}{58}%
\makeatletter
\providecommand \@ifxundefined [1]{%
 \@ifx{#1\undefined}
}%
\providecommand \@ifnum [1]{%
 \ifnum #1\expandafter \@firstoftwo
 \else \expandafter \@secondoftwo
 \fi
}%
\providecommand \@ifx [1]{%
 \ifx #1\expandafter \@firstoftwo
 \else \expandafter \@secondoftwo
 \fi
}%
\providecommand \natexlab [1]{#1}%
\providecommand \enquote  [1]{``#1''}%
\providecommand \bibnamefont  [1]{#1}%
\providecommand \bibfnamefont [1]{#1}%
\providecommand \citenamefont [1]{#1}%
\providecommand \href@noop [0]{\@secondoftwo}%
\providecommand \href [0]{\begingroup \@sanitize@url \@href}%
\providecommand \@href[1]{\@@startlink{#1}\@@href}%
\providecommand \@@href[1]{\endgroup#1\@@endlink}%
\providecommand \@sanitize@url [0]{\catcode `\\12\catcode `\$12\catcode
  `\&12\catcode `\#12\catcode `\^12\catcode `\_12\catcode `\%12\relax}%
\providecommand \@@startlink[1]{}%
\providecommand \@@endlink[0]{}%
\providecommand \url  [0]{\begingroup\@sanitize@url \@url }%
\providecommand \@url [1]{\endgroup\@href {#1}{\urlprefix }}%
\providecommand \urlprefix  [0]{URL }%
\providecommand \Eprint [0]{\href }%
\providecommand \doibase [0]{https://doi.org/}%
\providecommand \selectlanguage [0]{\@gobble}%
\providecommand \bibinfo  [0]{\@secondoftwo}%
\providecommand \bibfield  [0]{\@secondoftwo}%
\providecommand \translation [1]{[#1]}%
\providecommand \BibitemOpen [0]{}%
\providecommand \bibitemStop [0]{}%
\providecommand \bibitemNoStop [0]{.\EOS\space}%
\providecommand \EOS [0]{\spacefactor3000\relax}%
\providecommand \BibitemShut  [1]{\csname bibitem#1\endcsname}%
\let\auto@bib@innerbib\@empty
%</preamble>
\bibitem [{\citenamefont {Rubin}\ \emph {et~al.}(1982)\citenamefont {Rubin},
  \citenamefont {Ford}, \citenamefont {Thonnard},\ and\ \citenamefont
  {Burstein}}]{Rubin:1982kyu}%
  \BibitemOpen
  \bibfield  {author} {\bibinfo {author} {\bibfnamefont {V.~C.}\ \bibnamefont
  {Rubin}}, \bibinfo {author} {\bibfnamefont {W.~K.}\ \bibnamefont {Ford},
  \bibfnamefont {Jr.}}, \bibinfo {author} {\bibfnamefont {N.}~\bibnamefont
  {Thonnard}},\ and\ \bibinfo {author} {\bibfnamefont {D.}~\bibnamefont
  {Burstein}},\ }\bibfield  {title} {\bibinfo {title} {{Rotational properties
  of 23 SB galaxies}},\ }\href {https://doi.org/10.1086/160355} {\bibfield
  {journal} {\bibinfo  {journal} {Astrophys. J.}\ }\textbf {\bibinfo {volume}
  {261}},\ \bibinfo {pages} {439} (\bibinfo {year} {1982})}\BibitemShut
  {NoStop}%
\bibitem [{\citenamefont {Begeman}\ \emph {et~al.}(1991)\citenamefont
  {Begeman}, \citenamefont {Broeils},\ and\ \citenamefont
  {Sanders}}]{10.1093/mnras/249.3.523}%
  \BibitemOpen
  \bibfield  {author} {\bibinfo {author} {\bibfnamefont {K.~G.}\ \bibnamefont
  {Begeman}}, \bibinfo {author} {\bibfnamefont {A.~H.}\ \bibnamefont
  {Broeils}},\ and\ \bibinfo {author} {\bibfnamefont {R.~H.}\ \bibnamefont
  {Sanders}},\ }\bibfield  {title} {\bibinfo {title} {{Extended rotation curves
  of spiral galaxies: dark haloes and modified dynamics}},\ }\href
  {https://doi.org/10.1093/mnras/249.3.523} {\bibfield  {journal} {\bibinfo
  {journal} {Monthly Notices of the Royal Astronomical Society}\ }\textbf
  {\bibinfo {volume} {249}},\ \bibinfo {pages} {523} (\bibinfo {year}
  {1991})},\ \Eprint
  {https://arxiv.org/abs/https://academic.oup.com/mnras/article-pdf/249/3/523/18160929/mnras249-0523.pdf}
  {https://academic.oup.com/mnras/article-pdf/249/3/523/18160929/mnras249-0523.pdf}
  \BibitemShut {NoStop}%
\bibitem [{\citenamefont {Taylor}\ \emph {et~al.}(1998)\citenamefont {Taylor},
  \citenamefont {Dye}, \citenamefont {Broadhurst}, \citenamefont {Benitez},\
  and\ \citenamefont {van Kampen}}]{1998gravitational_lensing}%
  \BibitemOpen
  \bibfield  {author} {\bibinfo {author} {\bibfnamefont {A.~N.}\ \bibnamefont
  {Taylor}}, \bibinfo {author} {\bibfnamefont {S.}~\bibnamefont {Dye}},
  \bibinfo {author} {\bibfnamefont {T.~J.}\ \bibnamefont {Broadhurst}},
  \bibinfo {author} {\bibfnamefont {N.}~\bibnamefont {Benitez}},\ and\ \bibinfo
  {author} {\bibfnamefont {E.}~\bibnamefont {van Kampen}},\ }\bibfield  {title}
  {\bibinfo {title} {Gravitational lens magnification and the mass of abell
  1689},\ }\href {https://doi.org/10.1086/305827} {\bibfield  {journal}
  {\bibinfo  {journal} {The Astrophysical Journal}\ }\textbf {\bibinfo {volume}
  {501}},\ \bibinfo {pages} {539} (\bibinfo {year} {1998})}\BibitemShut
  {NoStop}%
\bibitem [{\citenamefont {Natarajan}\ \emph {et~al.}(2017)\citenamefont
  {Natarajan}, \citenamefont {Chadayammuri}, \citenamefont {Jauzac},
  \citenamefont {Richard}, \citenamefont {Kneib}, \citenamefont {Ebeling},
  \citenamefont {Jiang}, \citenamefont {van~den Bosch}, \citenamefont
  {Limousin}, \citenamefont {Jullo}, \citenamefont {Atek}, \citenamefont
  {Pillepich}, \citenamefont {Popa}, \citenamefont {Marinacci}, \citenamefont
  {Hernquist}, \citenamefont {Meneghetti},\ and\ \citenamefont
  {Vogelsberger}}]{10.1093/mnras/stw3385}%
  \BibitemOpen
  \bibfield  {author} {\bibinfo {author} {\bibfnamefont {P.}~\bibnamefont
  {Natarajan}}, \bibinfo {author} {\bibfnamefont {U.}~\bibnamefont
  {Chadayammuri}}, \bibinfo {author} {\bibfnamefont {M.}~\bibnamefont
  {Jauzac}}, \bibinfo {author} {\bibfnamefont {J.}~\bibnamefont {Richard}},
  \bibinfo {author} {\bibfnamefont {J.-P.}\ \bibnamefont {Kneib}}, \bibinfo
  {author} {\bibfnamefont {H.}~\bibnamefont {Ebeling}}, \bibinfo {author}
  {\bibfnamefont {F.}~\bibnamefont {Jiang}}, \bibinfo {author} {\bibfnamefont
  {F.}~\bibnamefont {van~den Bosch}}, \bibinfo {author} {\bibfnamefont
  {M.}~\bibnamefont {Limousin}}, \bibinfo {author} {\bibfnamefont
  {E.}~\bibnamefont {Jullo}}, \bibinfo {author} {\bibfnamefont
  {H.}~\bibnamefont {Atek}}, \bibinfo {author} {\bibfnamefont {A.}~\bibnamefont
  {Pillepich}}, \bibinfo {author} {\bibfnamefont {C.}~\bibnamefont {Popa}},
  \bibinfo {author} {\bibfnamefont {F.}~\bibnamefont {Marinacci}}, \bibinfo
  {author} {\bibfnamefont {L.}~\bibnamefont {Hernquist}}, \bibinfo {author}
  {\bibfnamefont {M.}~\bibnamefont {Meneghetti}},\ and\ \bibinfo {author}
  {\bibfnamefont {M.}~\bibnamefont {Vogelsberger}},\ }\bibfield  {title}
  {\bibinfo {title} {{Mapping substructure in the HST Frontier Fields cluster
  lenses and in cosmological simulations}},\ }\href
  {https://doi.org/10.1093/mnras/stw3385} {\bibfield  {journal} {\bibinfo
  {journal} {Monthly Notices of the Royal Astronomical Society}\ }\textbf
  {\bibinfo {volume} {468}},\ \bibinfo {pages} {1962} (\bibinfo {year}
  {2017})},\ \Eprint
  {https://arxiv.org/abs/https://academic.oup.com/mnras/article-pdf/468/2/1962/11210742/stw3385.pdf}
  {https://academic.oup.com/mnras/article-pdf/468/2/1962/11210742/stw3385.pdf}
  \BibitemShut {NoStop}%
\bibitem [{\citenamefont {Markevitch}\ \emph {et~al.}(2004)\citenamefont
  {Markevitch}, \citenamefont {Gonzalez}, \citenamefont {Clowe}, \citenamefont
  {Vikhlinin}, \citenamefont {Forman}, \citenamefont {Jones}, \citenamefont
  {Murray},\ and\ \citenamefont {Tucker}}]{Markevitch_2004}%
  \BibitemOpen
  \bibfield  {author} {\bibinfo {author} {\bibfnamefont {M.}~\bibnamefont
  {Markevitch}}, \bibinfo {author} {\bibfnamefont {A.~H.}\ \bibnamefont
  {Gonzalez}}, \bibinfo {author} {\bibfnamefont {D.}~\bibnamefont {Clowe}},
  \bibinfo {author} {\bibfnamefont {A.}~\bibnamefont {Vikhlinin}}, \bibinfo
  {author} {\bibfnamefont {W.}~\bibnamefont {Forman}}, \bibinfo {author}
  {\bibfnamefont {C.}~\bibnamefont {Jones}}, \bibinfo {author} {\bibfnamefont
  {S.}~\bibnamefont {Murray}},\ and\ \bibinfo {author} {\bibfnamefont
  {W.}~\bibnamefont {Tucker}},\ }\bibfield  {title} {\bibinfo {title} {Direct
  constraints on the dark matter self-interaction cross section from the
  merging galaxy cluster 1e 0657-56},\ }\href {https://doi.org/10.1086/383178}
  {\bibfield  {journal} {\bibinfo  {journal} {The Astrophysical Journal}\
  }\textbf {\bibinfo {volume} {606}},\ \bibinfo {pages} {819} (\bibinfo {year}
  {2004})}\BibitemShut {NoStop}%
\bibitem [{\citenamefont {Aghanim}\ \emph {et~al.}(2020)\citenamefont
  {Aghanim}, \citenamefont {Akrami}, \citenamefont {Ashdown}, \citenamefont
  {Aumont}, \citenamefont {Baccigalupi}, \citenamefont {Ballardini},
  \citenamefont {Banday}, \citenamefont {Barreiro}, \citenamefont {Bartolo},\
  and\ \citenamefont {et~al.}}]{2020Planck}%
  \BibitemOpen
  \bibfield  {author} {\bibinfo {author} {\bibfnamefont {N.}~\bibnamefont
  {Aghanim}}, \bibinfo {author} {\bibfnamefont {Y.}~\bibnamefont {Akrami}},
  \bibinfo {author} {\bibfnamefont {M.}~\bibnamefont {Ashdown}}, \bibinfo
  {author} {\bibfnamefont {J.}~\bibnamefont {Aumont}}, \bibinfo {author}
  {\bibfnamefont {C.}~\bibnamefont {Baccigalupi}}, \bibinfo {author}
  {\bibfnamefont {M.}~\bibnamefont {Ballardini}}, \bibinfo {author}
  {\bibfnamefont {A.~J.}\ \bibnamefont {Banday}}, \bibinfo {author}
  {\bibfnamefont {R.~B.}\ \bibnamefont {Barreiro}}, \bibinfo {author}
  {\bibfnamefont {N.}~\bibnamefont {Bartolo}},\ and\ \bibinfo {author}
  {\bibnamefont {et~al.}},\ }\bibfield  {title} {\bibinfo {title} {Planck 2018
  results},\ }\href {https://doi.org/10.1051/0004-6361/201833910} {\bibfield
  {journal} {\bibinfo  {journal} {Astronomy \& Astrophysics}\ }\textbf
  {\bibinfo {volume} {641}},\ \bibinfo {pages} {A6} (\bibinfo {year}
  {2020})}\BibitemShut {NoStop}%
\bibitem [{\citenamefont {Zyla}\ \emph {et~al.}(2020)\citenamefont {Zyla} \emph
  {et~al.}}]{Zyla:2020zbs}%
  \BibitemOpen
  \bibfield  {author} {\bibinfo {author} {\bibfnamefont {P.}~\bibnamefont
  {Zyla}} \emph {et~al.},\ }\bibfield  {title} {\bibinfo {title} {{Review of
  Particle Physics}},\ }\href {https://doi.org/10.1093/ptep/ptaa104} {\bibfield
   {journal} {\bibinfo  {journal} {PTEP}\ }\textbf {\bibinfo {volume} {2020}},\
  \bibinfo {pages} {083C01} (\bibinfo {year} {2020})}\BibitemShut {NoStop}%
\bibitem [{\citenamefont {Essig}\ \emph {et~al.}(2013)\citenamefont {Essig},
  \citenamefont {Jaros}, \citenamefont {Wester}, \citenamefont {Adrian},
  \citenamefont {Andreas}, \citenamefont {Averett}, \citenamefont {Baker},
  \citenamefont {Batell}, \citenamefont {Battaglieri}, \citenamefont {Beacham},
  \citenamefont {Beranek}, \citenamefont {Bjorken}, \citenamefont {Bossi},
  \citenamefont {Boyce}, \citenamefont {Cates}, \citenamefont {Celentano},
  \citenamefont {Chou}, \citenamefont {Cowan}, \citenamefont {Curciarello},
  \citenamefont {Davoudiasl}, \citenamefont {deNiverville}, \citenamefont
  {Vita}, \citenamefont {Denig}, \citenamefont {Dharmapalan}, \citenamefont
  {Dongwi}, \citenamefont {D{\"o}brich}, \citenamefont {Echenard},
  \citenamefont {Espriu}, \citenamefont {Fegan}, \citenamefont {Fisher},
  \citenamefont {Franklin}, \citenamefont {Gasparian}, \citenamefont
  {Gershtein}, \citenamefont {Graham}, \citenamefont {Graham}, \citenamefont
  {Haas}, \citenamefont {Hatzikoutelis}, \citenamefont {Holtrop}, \citenamefont
  {Irastorza}, \citenamefont {Izaguirre}, \citenamefont {Jaeckel},
  \citenamefont {Kahn}, \citenamefont {Kalantarians}, \citenamefont {Kohl},
  \citenamefont {Krnjaic}, \citenamefont {Kubarovsky}, \citenamefont {Lee},
  \citenamefont {Lindner}, \citenamefont {Lobanov}, \citenamefont {Marciano},
  \citenamefont {Marsh}, \citenamefont {Maruyama}, \citenamefont {McKeen},
  \citenamefont {Merkel}, \citenamefont {Moffeit}, \citenamefont {Monaghan},
  \citenamefont {Mueller}, \citenamefont {Nelson}, \citenamefont {Neil},
  \citenamefont {Oriunno}, \citenamefont {Pavlovic}, \citenamefont {Phillips},
  \citenamefont {Pivovaroff}, \citenamefont {Poltis}, \citenamefont {Pospelov},
  \citenamefont {Rajendran}, \citenamefont {Redondo}, \citenamefont {Ringwald},
  \citenamefont {Ritz}, \citenamefont {Ruz}, \citenamefont {Saenboonruang},
  \citenamefont {Schuster}, \citenamefont {Shinn}, \citenamefont {Slatyer},
  \citenamefont {Steffen}, \citenamefont {Stepanyan}, \citenamefont {Tanner},
  \citenamefont {Thaler}, \citenamefont {Tobar}, \citenamefont {Toro},
  \citenamefont {Upadye}, \citenamefont {de~Water}, \citenamefont {Vlahovic},
  \citenamefont {Vogel}, \citenamefont {Walker}, \citenamefont {Weltman},
  \citenamefont {Wojtsekhowski}, \citenamefont {Zhang},\ and\ \citenamefont
  {Zioutas}}]{essig2013dark}%
  \BibitemOpen
  \bibfield  {author} {\bibinfo {author} {\bibfnamefont {R.}~\bibnamefont
  {Essig}}, \bibinfo {author} {\bibfnamefont {J.~A.}\ \bibnamefont {Jaros}},
  \bibinfo {author} {\bibfnamefont {W.}~\bibnamefont {Wester}}, \bibinfo
  {author} {\bibfnamefont {P.~H.}\ \bibnamefont {Adrian}}, \bibinfo {author}
  {\bibfnamefont {S.}~\bibnamefont {Andreas}}, \bibinfo {author} {\bibfnamefont
  {T.}~\bibnamefont {Averett}}, \bibinfo {author} {\bibfnamefont
  {O.}~\bibnamefont {Baker}}, \bibinfo {author} {\bibfnamefont
  {B.}~\bibnamefont {Batell}}, \bibinfo {author} {\bibfnamefont
  {M.}~\bibnamefont {Battaglieri}}, \bibinfo {author} {\bibfnamefont
  {J.}~\bibnamefont {Beacham}}, \bibinfo {author} {\bibfnamefont
  {T.}~\bibnamefont {Beranek}}, \bibinfo {author} {\bibfnamefont {J.~D.}\
  \bibnamefont {Bjorken}}, \bibinfo {author} {\bibfnamefont {F.}~\bibnamefont
  {Bossi}}, \bibinfo {author} {\bibfnamefont {J.~R.}\ \bibnamefont {Boyce}},
  \bibinfo {author} {\bibfnamefont {G.~D.}\ \bibnamefont {Cates}}, \bibinfo
  {author} {\bibfnamefont {A.}~\bibnamefont {Celentano}}, \bibinfo {author}
  {\bibfnamefont {A.~S.}\ \bibnamefont {Chou}}, \bibinfo {author}
  {\bibfnamefont {R.}~\bibnamefont {Cowan}}, \bibinfo {author} {\bibfnamefont
  {F.}~\bibnamefont {Curciarello}}, \bibinfo {author} {\bibfnamefont
  {H.}~\bibnamefont {Davoudiasl}}, \bibinfo {author} {\bibfnamefont
  {P.}~\bibnamefont {deNiverville}}, \bibinfo {author} {\bibfnamefont {R.~D.}\
  \bibnamefont {Vita}}, \bibinfo {author} {\bibfnamefont {A.}~\bibnamefont
  {Denig}}, \bibinfo {author} {\bibfnamefont {R.}~\bibnamefont {Dharmapalan}},
  \bibinfo {author} {\bibfnamefont {B.}~\bibnamefont {Dongwi}}, \bibinfo
  {author} {\bibfnamefont {B.}~\bibnamefont {D{\"o}brich}}, \bibinfo {author}
  {\bibfnamefont {B.}~\bibnamefont {Echenard}}, \bibinfo {author}
  {\bibfnamefont {D.}~\bibnamefont {Espriu}}, \bibinfo {author} {\bibfnamefont
  {S.}~\bibnamefont {Fegan}}, \bibinfo {author} {\bibfnamefont
  {P.}~\bibnamefont {Fisher}}, \bibinfo {author} {\bibfnamefont {G.~B.}\
  \bibnamefont {Franklin}}, \bibinfo {author} {\bibfnamefont {A.}~\bibnamefont
  {Gasparian}}, \bibinfo {author} {\bibfnamefont {Y.}~\bibnamefont
  {Gershtein}}, \bibinfo {author} {\bibfnamefont {M.}~\bibnamefont {Graham}},
  \bibinfo {author} {\bibfnamefont {P.~W.}\ \bibnamefont {Graham}}, \bibinfo
  {author} {\bibfnamefont {A.}~\bibnamefont {Haas}}, \bibinfo {author}
  {\bibfnamefont {A.}~\bibnamefont {Hatzikoutelis}}, \bibinfo {author}
  {\bibfnamefont {M.}~\bibnamefont {Holtrop}}, \bibinfo {author} {\bibfnamefont
  {I.}~\bibnamefont {Irastorza}}, \bibinfo {author} {\bibfnamefont
  {E.}~\bibnamefont {Izaguirre}}, \bibinfo {author} {\bibfnamefont
  {J.}~\bibnamefont {Jaeckel}}, \bibinfo {author} {\bibfnamefont
  {Y.}~\bibnamefont {Kahn}}, \bibinfo {author} {\bibfnamefont {N.}~\bibnamefont
  {Kalantarians}}, \bibinfo {author} {\bibfnamefont {M.}~\bibnamefont {Kohl}},
  \bibinfo {author} {\bibfnamefont {G.}~\bibnamefont {Krnjaic}}, \bibinfo
  {author} {\bibfnamefont {V.}~\bibnamefont {Kubarovsky}}, \bibinfo {author}
  {\bibfnamefont {H.-S.}\ \bibnamefont {Lee}}, \bibinfo {author} {\bibfnamefont
  {A.}~\bibnamefont {Lindner}}, \bibinfo {author} {\bibfnamefont
  {A.}~\bibnamefont {Lobanov}}, \bibinfo {author} {\bibfnamefont {W.~J.}\
  \bibnamefont {Marciano}}, \bibinfo {author} {\bibfnamefont {D.~J.~E.}\
  \bibnamefont {Marsh}}, \bibinfo {author} {\bibfnamefont {T.}~\bibnamefont
  {Maruyama}}, \bibinfo {author} {\bibfnamefont {D.}~\bibnamefont {McKeen}},
  \bibinfo {author} {\bibfnamefont {H.}~\bibnamefont {Merkel}}, \bibinfo
  {author} {\bibfnamefont {K.}~\bibnamefont {Moffeit}}, \bibinfo {author}
  {\bibfnamefont {P.}~\bibnamefont {Monaghan}}, \bibinfo {author}
  {\bibfnamefont {G.}~\bibnamefont {Mueller}}, \bibinfo {author} {\bibfnamefont
  {T.~K.}\ \bibnamefont {Nelson}}, \bibinfo {author} {\bibfnamefont {G.~R.}\
  \bibnamefont {Neil}}, \bibinfo {author} {\bibfnamefont {M.}~\bibnamefont
  {Oriunno}}, \bibinfo {author} {\bibfnamefont {Z.}~\bibnamefont {Pavlovic}},
  \bibinfo {author} {\bibfnamefont {S.~K.}\ \bibnamefont {Phillips}}, \bibinfo
  {author} {\bibfnamefont {M.~J.}\ \bibnamefont {Pivovaroff}}, \bibinfo
  {author} {\bibfnamefont {R.}~\bibnamefont {Poltis}}, \bibinfo {author}
  {\bibfnamefont {M.}~\bibnamefont {Pospelov}}, \bibinfo {author}
  {\bibfnamefont {S.}~\bibnamefont {Rajendran}}, \bibinfo {author}
  {\bibfnamefont {J.}~\bibnamefont {Redondo}}, \bibinfo {author} {\bibfnamefont
  {A.}~\bibnamefont {Ringwald}}, \bibinfo {author} {\bibfnamefont
  {A.}~\bibnamefont {Ritz}}, \bibinfo {author} {\bibfnamefont {J.}~\bibnamefont
  {Ruz}}, \bibinfo {author} {\bibfnamefont {K.}~\bibnamefont {Saenboonruang}},
  \bibinfo {author} {\bibfnamefont {P.}~\bibnamefont {Schuster}}, \bibinfo
  {author} {\bibfnamefont {M.}~\bibnamefont {Shinn}}, \bibinfo {author}
  {\bibfnamefont {T.~R.}\ \bibnamefont {Slatyer}}, \bibinfo {author}
  {\bibfnamefont {J.~H.}\ \bibnamefont {Steffen}}, \bibinfo {author}
  {\bibfnamefont {S.}~\bibnamefont {Stepanyan}}, \bibinfo {author}
  {\bibfnamefont {D.~B.}\ \bibnamefont {Tanner}}, \bibinfo {author}
  {\bibfnamefont {J.}~\bibnamefont {Thaler}}, \bibinfo {author} {\bibfnamefont
  {M.~E.}\ \bibnamefont {Tobar}}, \bibinfo {author} {\bibfnamefont
  {N.}~\bibnamefont {Toro}}, \bibinfo {author} {\bibfnamefont {A.}~\bibnamefont
  {Upadye}}, \bibinfo {author} {\bibfnamefont {R.~V.}\ \bibnamefont
  {de~Water}}, \bibinfo {author} {\bibfnamefont {B.}~\bibnamefont {Vlahovic}},
  \bibinfo {author} {\bibfnamefont {J.~K.}\ \bibnamefont {Vogel}}, \bibinfo
  {author} {\bibfnamefont {D.}~\bibnamefont {Walker}}, \bibinfo {author}
  {\bibfnamefont {A.}~\bibnamefont {Weltman}}, \bibinfo {author} {\bibfnamefont
  {B.}~\bibnamefont {Wojtsekhowski}}, \bibinfo {author} {\bibfnamefont
  {S.}~\bibnamefont {Zhang}},\ and\ \bibinfo {author} {\bibfnamefont
  {K.}~\bibnamefont {Zioutas}},\ }\href@noop {} {\bibinfo {title} {Dark sectors
  and new, light, weakly-coupled particles}} (\bibinfo {year} {2013}),\ \Eprint
  {https://arxiv.org/abs/1311.0029} {arXiv:1311.0029 [hep-ph]} \BibitemShut
  {NoStop}%
\bibitem [{\citenamefont {Ghosh}\ \emph {et~al.}(2021)\citenamefont {Ghosh},
  \citenamefont {Ruddy}, \citenamefont {Jewell}, \citenamefont {Leder},\ and\
  \citenamefont {Maruyama}}]{PhysRevD.104.092016}%
  \BibitemOpen
  \bibfield  {author} {\bibinfo {author} {\bibfnamefont {S.}~\bibnamefont
  {Ghosh}}, \bibinfo {author} {\bibfnamefont {E.~P.}\ \bibnamefont {Ruddy}},
  \bibinfo {author} {\bibfnamefont {M.~J.}\ \bibnamefont {Jewell}}, \bibinfo
  {author} {\bibfnamefont {A.~F.}\ \bibnamefont {Leder}},\ and\ \bibinfo
  {author} {\bibfnamefont {R.~H.}\ \bibnamefont {Maruyama}},\ }\bibfield
  {title} {\bibinfo {title} {Searching for dark photons with existing haloscope
  data},\ }\href {https://doi.org/10.1103/PhysRevD.104.092016} {\bibfield
  {journal} {\bibinfo  {journal} {Phys. Rev. D}\ }\textbf {\bibinfo {volume}
  {104}},\ \bibinfo {pages} {092016} (\bibinfo {year} {2021})}\BibitemShut
  {NoStop}%
\bibitem [{\citenamefont {Caputo}\ \emph {et~al.}(2021)\citenamefont {Caputo},
  \citenamefont {Millar}, \citenamefont {O'Hare},\ and\ \citenamefont
  {Vitagliano}}]{PhysRevD.104.095029}%
  \BibitemOpen
  \bibfield  {author} {\bibinfo {author} {\bibfnamefont {A.}~\bibnamefont
  {Caputo}}, \bibinfo {author} {\bibfnamefont {A.~J.}\ \bibnamefont {Millar}},
  \bibinfo {author} {\bibfnamefont {C.~A.~J.}\ \bibnamefont {O'Hare}},\ and\
  \bibinfo {author} {\bibfnamefont {E.}~\bibnamefont {Vitagliano}},\ }\bibfield
   {title} {\bibinfo {title} {Dark photon limits: A handbook},\ }\href
  {https://doi.org/10.1103/PhysRevD.104.095029} {\bibfield  {journal} {\bibinfo
   {journal} {Phys. Rev. D}\ }\textbf {\bibinfo {volume} {104}},\ \bibinfo
  {pages} {095029} (\bibinfo {year} {2021})}\BibitemShut {NoStop}%
\bibitem [{\citenamefont {Holdom}(1986{\natexlab{a}})}]{HOLDOM198665}%
  \BibitemOpen
  \bibfield  {author} {\bibinfo {author} {\bibfnamefont {B.}~\bibnamefont
  {Holdom}},\ }\bibfield  {title} {\bibinfo {title} {Searching for $\epsilon$
  charges and a new u(1)},\ }\href
  {https://doi.org/https://doi.org/10.1016/0370-2693(86)90470-3} {\bibfield
  {journal} {\bibinfo  {journal} {Physics Letters B}\ }\textbf {\bibinfo
  {volume} {178}},\ \bibinfo {pages} {65} (\bibinfo {year}
  {1986}{\natexlab{a}})}\BibitemShut {NoStop}%
\bibitem [{\citenamefont {Holdom}(1986{\natexlab{b}})}]{HOLDOM1986196}%
  \BibitemOpen
  \bibfield  {author} {\bibinfo {author} {\bibfnamefont {B.}~\bibnamefont
  {Holdom}},\ }\bibfield  {title} {\bibinfo {title} {Two u(1)'s and $\epsilon$
  charge shifts},\ }\href
  {https://doi.org/https://doi.org/10.1016/0370-2693(86)91377-8} {\bibfield
  {journal} {\bibinfo  {journal} {Physics Letters B}\ }\textbf {\bibinfo
  {volume} {166}},\ \bibinfo {pages} {196} (\bibinfo {year}
  {1986}{\natexlab{b}})}\BibitemShut {NoStop}%
\bibitem [{\citenamefont {Pospelov}\ \emph {et~al.}(2008)\citenamefont
  {Pospelov}, \citenamefont {Ritz},\ and\ \citenamefont
  {Voloshin}}]{PhysRevD.78.115012}%
  \BibitemOpen
  \bibfield  {author} {\bibinfo {author} {\bibfnamefont {M.}~\bibnamefont
  {Pospelov}}, \bibinfo {author} {\bibfnamefont {A.}~\bibnamefont {Ritz}},\
  and\ \bibinfo {author} {\bibfnamefont {M.}~\bibnamefont {Voloshin}},\
  }\bibfield  {title} {\bibinfo {title} {Bosonic super-wimps as kev-scale dark
  matter},\ }\href {https://doi.org/10.1103/PhysRevD.78.115012} {\bibfield
  {journal} {\bibinfo  {journal} {Phys. Rev. D}\ }\textbf {\bibinfo {volume}
  {78}},\ \bibinfo {pages} {115012} (\bibinfo {year} {2008})}\BibitemShut
  {NoStop}%
\bibitem [{\citenamefont {Turner}(1990)}]{PhysRevD.42.3572}%
  \BibitemOpen
  \bibfield  {author} {\bibinfo {author} {\bibfnamefont {M.~S.}\ \bibnamefont
  {Turner}},\ }\bibfield  {title} {\bibinfo {title} {Periodic signatures for
  the detection of cosmic axions},\ }\href
  {https://doi.org/10.1103/PhysRevD.42.3572} {\bibfield  {journal} {\bibinfo
  {journal} {Phys. Rev. D}\ }\textbf {\bibinfo {volume} {42}},\ \bibinfo
  {pages} {3572} (\bibinfo {year} {1990})}\BibitemShut {NoStop}%
\bibitem [{\citenamefont {Jimenez}\ \emph {et~al.}(2003)\citenamefont
  {Jimenez}, \citenamefont {Verde},\ and\ \citenamefont
  {Oh}}]{10.1046/j.1365-8711.2003.06165.x}%
  \BibitemOpen
  \bibfield  {author} {\bibinfo {author} {\bibfnamefont {R.}~\bibnamefont
  {Jimenez}}, \bibinfo {author} {\bibfnamefont {L.}~\bibnamefont {Verde}},\
  and\ \bibinfo {author} {\bibfnamefont {S.~P.}\ \bibnamefont {Oh}},\
  }\bibfield  {title} {\bibinfo {title} {{Dark halo properties from rotation
  curves}},\ }\href {https://doi.org/10.1046/j.1365-8711.2003.06165.x}
  {\bibfield  {journal} {\bibinfo  {journal} {Monthly Notices of the Royal
  Astronomical Society}\ }\textbf {\bibinfo {volume} {339}},\ \bibinfo {pages}
  {243} (\bibinfo {year} {2003})},\ \Eprint
  {https://arxiv.org/abs/https://academic.oup.com/mnras/article-pdf/339/1/243/3557948/339-1-243.pdf}
  {https://academic.oup.com/mnras/article-pdf/339/1/243/3557948/339-1-243.pdf}
  \BibitemShut {NoStop}%
\bibitem [{\citenamefont {Graham}\ \emph {et~al.}(2016)\citenamefont {Graham},
  \citenamefont {Mardon},\ and\ \citenamefont
  {Rajendran}}]{PhysRevD.93.103520}%
  \BibitemOpen
  \bibfield  {author} {\bibinfo {author} {\bibfnamefont {P.~W.}\ \bibnamefont
  {Graham}}, \bibinfo {author} {\bibfnamefont {J.}~\bibnamefont {Mardon}},\
  and\ \bibinfo {author} {\bibfnamefont {S.}~\bibnamefont {Rajendran}},\
  }\bibfield  {title} {\bibinfo {title} {Vector dark matter from inflationary
  fluctuations},\ }\href {https://doi.org/10.1103/PhysRevD.93.103520}
  {\bibfield  {journal} {\bibinfo  {journal} {Phys. Rev. D}\ }\textbf {\bibinfo
  {volume} {93}},\ \bibinfo {pages} {103520} (\bibinfo {year}
  {2016})}\BibitemShut {NoStop}%
\bibitem [{\citenamefont {Arias}\ \emph {et~al.}(2012)\citenamefont {Arias},
  \citenamefont {Cadamuro}, \citenamefont {Goodsell}, \citenamefont {Jaeckel},
  \citenamefont {Redondo},\ and\ \citenamefont {Ringwald}}]{Arias_2012}%
  \BibitemOpen
  \bibfield  {author} {\bibinfo {author} {\bibfnamefont {P.}~\bibnamefont
  {Arias}}, \bibinfo {author} {\bibfnamefont {D.}~\bibnamefont {Cadamuro}},
  \bibinfo {author} {\bibfnamefont {M.}~\bibnamefont {Goodsell}}, \bibinfo
  {author} {\bibfnamefont {J.}~\bibnamefont {Jaeckel}}, \bibinfo {author}
  {\bibfnamefont {J.}~\bibnamefont {Redondo}},\ and\ \bibinfo {author}
  {\bibfnamefont {A.}~\bibnamefont {Ringwald}},\ }\bibfield  {title} {\bibinfo
  {title} {{WISPy} cold dark matter},\ }\href
  {https://doi.org/10.1088/1475-7516/2012/06/013} {\bibfield  {journal}
  {\bibinfo  {journal} {Journal of Cosmology and Astroparticle Physics}\
  }\textbf {\bibinfo {volume} {2012}}\bibinfo  {number} { (06)},\ \bibinfo
  {pages} {013}}\BibitemShut {NoStop}%
\bibitem [{\citenamefont {Sikivie}(1983)}]{PhysRevLett.51.1415}%
  \BibitemOpen
\bibfield  {number} {  }\bibfield  {author} {\bibinfo {author} {\bibfnamefont
  {P.}~\bibnamefont {Sikivie}},\ }\bibfield  {title} {\bibinfo {title}
  {Experimental tests of the ``invisible'' axion},\ }\href
  {https://doi.org/10.1103/PhysRevLett.51.1415} {\bibfield  {journal} {\bibinfo
   {journal} {Phys. Rev. Lett.}\ }\textbf {\bibinfo {volume} {51}},\ \bibinfo
  {pages} {1415} (\bibinfo {year} {1983})}\BibitemShut {NoStop}%
\bibitem [{\citenamefont {Cervantes}\ \emph
  {et~al.}(2022{\natexlab{a}})\citenamefont {Cervantes}, \citenamefont
  {Carosi}, \citenamefont {Kimes}, \citenamefont {Hanretty}, \citenamefont
  {LaRoque}, \citenamefont {Leum}, \citenamefont {Mohapatra}, \citenamefont
  {Oblath}, \citenamefont {Ottens}, \citenamefont {Park}, \citenamefont
  {Rybka}, \citenamefont {Sinnis},\ and\ \citenamefont
  {Yang}}]{PhysRevD.106.102002}%
  \BibitemOpen
  \bibfield  {author} {\bibinfo {author} {\bibfnamefont {R.}~\bibnamefont
  {Cervantes}}, \bibinfo {author} {\bibfnamefont {G.}~\bibnamefont {Carosi}},
  \bibinfo {author} {\bibfnamefont {S.}~\bibnamefont {Kimes}}, \bibinfo
  {author} {\bibfnamefont {C.}~\bibnamefont {Hanretty}}, \bibinfo {author}
  {\bibfnamefont {B.~H.}\ \bibnamefont {LaRoque}}, \bibinfo {author}
  {\bibfnamefont {G.}~\bibnamefont {Leum}}, \bibinfo {author} {\bibfnamefont
  {P.}~\bibnamefont {Mohapatra}}, \bibinfo {author} {\bibfnamefont {N.~S.}\
  \bibnamefont {Oblath}}, \bibinfo {author} {\bibfnamefont {R.}~\bibnamefont
  {Ottens}}, \bibinfo {author} {\bibfnamefont {Y.}~\bibnamefont {Park}},
  \bibinfo {author} {\bibfnamefont {G.}~\bibnamefont {Rybka}}, \bibinfo
  {author} {\bibfnamefont {J.}~\bibnamefont {Sinnis}},\ and\ \bibinfo {author}
  {\bibfnamefont {J.}~\bibnamefont {Yang}},\ }\bibfield  {title} {\bibinfo
  {title} {Admx-orpheus first search for $70\text{ }\text{
  }\ensuremath{\mu}\mathrm{eV}$ dark photon dark matter: Detailed design,
  operations, and analysis},\ }\href
  {https://doi.org/10.1103/PhysRevD.106.102002} {\bibfield  {journal} {\bibinfo
   {journal} {Phys. Rev. D}\ }\textbf {\bibinfo {volume} {106}},\ \bibinfo
  {pages} {102002} (\bibinfo {year} {2022}{\natexlab{a}})}\BibitemShut
  {NoStop}%
\bibitem [{\citenamefont {Kim}\ \emph {et~al.}(2020)\citenamefont {Kim},
  \citenamefont {Jeong}, \citenamefont {Youn}, \citenamefont {Kim},\ and\
  \citenamefont {Semertzidis}}]{Kim_2020}%
  \BibitemOpen
  \bibfield  {author} {\bibinfo {author} {\bibfnamefont {D.}~\bibnamefont
  {Kim}}, \bibinfo {author} {\bibfnamefont {J.}~\bibnamefont {Jeong}}, \bibinfo
  {author} {\bibfnamefont {S.}~\bibnamefont {Youn}}, \bibinfo {author}
  {\bibfnamefont {Y.}~\bibnamefont {Kim}},\ and\ \bibinfo {author}
  {\bibfnamefont {Y.~K.}\ \bibnamefont {Semertzidis}},\ }\bibfield  {title}
  {\bibinfo {title} {Revisiting the detection rate for axion haloscopes},\
  }\href {https://doi.org/10.1088/1475-7516/2020/03/066} {\bibfield  {journal}
  {\bibinfo  {journal} {Journal of Cosmology and Astroparticle Physics}\
  }\textbf {\bibinfo {volume} {2020}}\bibinfo  {number} { (03)},\ \bibinfo
  {pages} {066}}\BibitemShut {NoStop}%
\bibitem [{\citenamefont {Sikivie}(1985)}]{PhysRevD.32.2988}%
  \BibitemOpen
\bibfield  {number} {  }\bibfield  {author} {\bibinfo {author} {\bibfnamefont
  {P.}~\bibnamefont {Sikivie}},\ }\bibfield  {title} {\bibinfo {title}
  {Detection rates for ``invisible''-axion searches},\ }\href
  {https://doi.org/10.1103/PhysRevD.32.2988} {\bibfield  {journal} {\bibinfo
  {journal} {Phys. Rev. D}\ }\textbf {\bibinfo {volume} {32}},\ \bibinfo
  {pages} {2988} (\bibinfo {year} {1985})}\BibitemShut {NoStop}%
\bibitem [{\citenamefont {Romanenko}\ \emph {et~al.}(2020)\citenamefont
  {Romanenko}, \citenamefont {Pilipenko}, \citenamefont {Zorzetti},
  \citenamefont {Frolov}, \citenamefont {Awida}, \citenamefont {Belomestnykh},
  \citenamefont {Posen},\ and\ \citenamefont
  {Grassellino}}]{PhysRevApplied.13.034032}%
  \BibitemOpen
  \bibfield  {author} {\bibinfo {author} {\bibfnamefont {A.}~\bibnamefont
  {Romanenko}}, \bibinfo {author} {\bibfnamefont {R.}~\bibnamefont
  {Pilipenko}}, \bibinfo {author} {\bibfnamefont {S.}~\bibnamefont {Zorzetti}},
  \bibinfo {author} {\bibfnamefont {D.}~\bibnamefont {Frolov}}, \bibinfo
  {author} {\bibfnamefont {M.}~\bibnamefont {Awida}}, \bibinfo {author}
  {\bibfnamefont {S.}~\bibnamefont {Belomestnykh}}, \bibinfo {author}
  {\bibfnamefont {S.}~\bibnamefont {Posen}},\ and\ \bibinfo {author}
  {\bibfnamefont {A.}~\bibnamefont {Grassellino}},\ }\bibfield  {title}
  {\bibinfo {title} {Three-dimensional superconducting resonators at $t=20$ mk
  with photon lifetimes up to $\ensuremath{\tau}=2$ s},\ }\href
  {https://doi.org/10.1103/PhysRevApplied.13.034032} {\bibfield  {journal}
  {\bibinfo  {journal} {Phys. Rev. Applied}\ }\textbf {\bibinfo {volume}
  {13}},\ \bibinfo {pages} {034032} (\bibinfo {year} {2020})}\BibitemShut
  {NoStop}%
\bibitem [{\citenamefont {Alesini}\ \emph {et~al.}(2021)\citenamefont
  {Alesini}, \citenamefont {Braggio}, \citenamefont {Carugno}, \citenamefont
  {Crescini}, \citenamefont {{D’ Agostino}}, \citenamefont {{Di Gioacchino}},
  \citenamefont {{Di Vora}}, \citenamefont {Falferi}, \citenamefont
  {Gambardella}, \citenamefont {Gatti}, \citenamefont {Iannone}, \citenamefont
  {Ligi}, \citenamefont {Lombardi}, \citenamefont {Maccarrone}, \citenamefont
  {Ortolan}, \citenamefont {Pengo}, \citenamefont {Pira}, \citenamefont
  {Rettaroli}, \citenamefont {Ruoso}, \citenamefont {Taffarello},\ and\
  \citenamefont {Tocci}}]{ALESINI2021164641}%
  \BibitemOpen
  \bibfield  {author} {\bibinfo {author} {\bibfnamefont {D.}~\bibnamefont
  {Alesini}}, \bibinfo {author} {\bibfnamefont {C.}~\bibnamefont {Braggio}},
  \bibinfo {author} {\bibfnamefont {G.}~\bibnamefont {Carugno}}, \bibinfo
  {author} {\bibfnamefont {N.}~\bibnamefont {Crescini}}, \bibinfo {author}
  {\bibfnamefont {D.}~\bibnamefont {{D’ Agostino}}}, \bibinfo {author}
  {\bibfnamefont {D.}~\bibnamefont {{Di Gioacchino}}}, \bibinfo {author}
  {\bibfnamefont {R.}~\bibnamefont {{Di Vora}}}, \bibinfo {author}
  {\bibfnamefont {P.}~\bibnamefont {Falferi}}, \bibinfo {author} {\bibfnamefont
  {U.}~\bibnamefont {Gambardella}}, \bibinfo {author} {\bibfnamefont
  {C.}~\bibnamefont {Gatti}}, \bibinfo {author} {\bibfnamefont
  {G.}~\bibnamefont {Iannone}}, \bibinfo {author} {\bibfnamefont
  {C.}~\bibnamefont {Ligi}}, \bibinfo {author} {\bibfnamefont {A.}~\bibnamefont
  {Lombardi}}, \bibinfo {author} {\bibfnamefont {G.}~\bibnamefont
  {Maccarrone}}, \bibinfo {author} {\bibfnamefont {A.}~\bibnamefont {Ortolan}},
  \bibinfo {author} {\bibfnamefont {R.}~\bibnamefont {Pengo}}, \bibinfo
  {author} {\bibfnamefont {C.}~\bibnamefont {Pira}}, \bibinfo {author}
  {\bibfnamefont {A.}~\bibnamefont {Rettaroli}}, \bibinfo {author}
  {\bibfnamefont {G.}~\bibnamefont {Ruoso}}, \bibinfo {author} {\bibfnamefont
  {L.}~\bibnamefont {Taffarello}},\ and\ \bibinfo {author} {\bibfnamefont
  {S.}~\bibnamefont {Tocci}},\ }\bibfield  {title} {\bibinfo {title}
  {Realization of a high quality factor resonator with hollow dielectric
  cylinders for axion searches},\ }\href
  {https://doi.org/https://doi.org/10.1016/j.nima.2020.164641} {\bibfield
  {journal} {\bibinfo  {journal} {Nuclear Instruments and Methods in Physics
  Research Section A: Accelerators, Spectrometers, Detectors and Associated
  Equipment}\ }\textbf {\bibinfo {volume} {985}},\ \bibinfo {pages} {164641}
  (\bibinfo {year} {2021})}\BibitemShut {NoStop}%
\bibitem [{\citenamefont {Posen}\ \emph {et~al.}(2022)\citenamefont {Posen},
  \citenamefont {Checchin}, \citenamefont {Melnychuk}, \citenamefont {Ring},\
  and\ \citenamefont {Gonin}}]{https://doi.org/10.48550/arxiv.2201.10733}%
  \BibitemOpen
  \bibfield  {author} {\bibinfo {author} {\bibfnamefont {S.}~\bibnamefont
  {Posen}}, \bibinfo {author} {\bibfnamefont {M.}~\bibnamefont {Checchin}},
  \bibinfo {author} {\bibfnamefont {O.~S.}\ \bibnamefont {Melnychuk}}, \bibinfo
  {author} {\bibfnamefont {T.}~\bibnamefont {Ring}},\ and\ \bibinfo {author}
  {\bibfnamefont {I.}~\bibnamefont {Gonin}},\ }\href
  {https://doi.org/10.48550/ARXIV.2201.10733} {\bibinfo {title} {Measurement of
  high quality factor superconducting cavities in tesla-scale magnetic fields
  for dark matter searches}} (\bibinfo {year} {2022})\BibitemShut {NoStop}%
\bibitem [{\citenamefont {Di~Vora}\ \emph {et~al.}(2022)\citenamefont
  {Di~Vora}, \citenamefont {Alesini}, \citenamefont {Braggio}, \citenamefont
  {Carugno}, \citenamefont {Crescini}, \citenamefont {D'Agostino},
  \citenamefont {Di~Gioacchino}, \citenamefont {Falferi}, \citenamefont
  {Gambardella}, \citenamefont {Gatti}, \citenamefont {Iannone}, \citenamefont
  {Ligi}, \citenamefont {Lombardi}, \citenamefont {Maccarrone}, \citenamefont
  {Ortolan}, \citenamefont {Pengo}, \citenamefont {Rettaroli}, \citenamefont
  {Ruoso}, \citenamefont {Taffarello},\ and\ \citenamefont
  {Tocci}}]{PhysRevApplied.17.054013}%
  \BibitemOpen
  \bibfield  {author} {\bibinfo {author} {\bibfnamefont {R.}~\bibnamefont
  {Di~Vora}}, \bibinfo {author} {\bibfnamefont {D.}~\bibnamefont {Alesini}},
  \bibinfo {author} {\bibfnamefont {C.}~\bibnamefont {Braggio}}, \bibinfo
  {author} {\bibfnamefont {G.}~\bibnamefont {Carugno}}, \bibinfo {author}
  {\bibfnamefont {N.}~\bibnamefont {Crescini}}, \bibinfo {author}
  {\bibfnamefont {D.}~\bibnamefont {D'Agostino}}, \bibinfo {author}
  {\bibfnamefont {D.}~\bibnamefont {Di~Gioacchino}}, \bibinfo {author}
  {\bibfnamefont {P.}~\bibnamefont {Falferi}}, \bibinfo {author} {\bibfnamefont
  {U.}~\bibnamefont {Gambardella}}, \bibinfo {author} {\bibfnamefont
  {C.}~\bibnamefont {Gatti}}, \bibinfo {author} {\bibfnamefont
  {G.}~\bibnamefont {Iannone}}, \bibinfo {author} {\bibfnamefont
  {C.}~\bibnamefont {Ligi}}, \bibinfo {author} {\bibfnamefont {A.}~\bibnamefont
  {Lombardi}}, \bibinfo {author} {\bibfnamefont {G.}~\bibnamefont
  {Maccarrone}}, \bibinfo {author} {\bibfnamefont {A.}~\bibnamefont {Ortolan}},
  \bibinfo {author} {\bibfnamefont {R.}~\bibnamefont {Pengo}}, \bibinfo
  {author} {\bibfnamefont {A.}~\bibnamefont {Rettaroli}}, \bibinfo {author}
  {\bibfnamefont {G.}~\bibnamefont {Ruoso}}, \bibinfo {author} {\bibfnamefont
  {L.}~\bibnamefont {Taffarello}},\ and\ \bibinfo {author} {\bibfnamefont
  {S.}~\bibnamefont {Tocci}},\ }\bibfield  {title} {\bibinfo {title} {High-$q$
  microwave dielectric resonator for axion dark-matter haloscopes},\ }\href
  {https://doi.org/10.1103/PhysRevApplied.17.054013} {\bibfield  {journal}
  {\bibinfo  {journal} {Phys. Rev. Applied}\ }\textbf {\bibinfo {volume}
  {17}},\ \bibinfo {pages} {054013} (\bibinfo {year} {2022})}\BibitemShut
  {NoStop}%
\bibitem [{\citenamefont {Ahn}\ \emph {et~al.}(2022)\citenamefont {Ahn},
  \citenamefont {Kwon}, \citenamefont {Chung}, \citenamefont {Jang},
  \citenamefont {Lee}, \citenamefont {Lee}, \citenamefont {Youn}, \citenamefont
  {Byun}, \citenamefont {Youm},\ and\ \citenamefont
  {Semertzidis}}]{PhysRevApplied.17.L061005}%
  \BibitemOpen
  \bibfield  {author} {\bibinfo {author} {\bibfnamefont {D.}~\bibnamefont
  {Ahn}}, \bibinfo {author} {\bibfnamefont {O.}~\bibnamefont {Kwon}}, \bibinfo
  {author} {\bibfnamefont {W.}~\bibnamefont {Chung}}, \bibinfo {author}
  {\bibfnamefont {W.}~\bibnamefont {Jang}}, \bibinfo {author} {\bibfnamefont
  {D.}~\bibnamefont {Lee}}, \bibinfo {author} {\bibfnamefont {J.}~\bibnamefont
  {Lee}}, \bibinfo {author} {\bibfnamefont {S.~W.}\ \bibnamefont {Youn}},
  \bibinfo {author} {\bibfnamefont {H.}~\bibnamefont {Byun}}, \bibinfo {author}
  {\bibfnamefont {D.}~\bibnamefont {Youm}},\ and\ \bibinfo {author}
  {\bibfnamefont {Y.~K.}\ \bibnamefont {Semertzidis}},\ }\bibfield  {title}
  {\bibinfo {title} {Biaxially textured
  ${\mathrm{yba}}_{2}{\mathrm{cu}}_{3}{\mathrm{o}}_{7\ensuremath{-}x}$
  microwave cavity in a high magnetic field for a dark-matter axion search},\
  }\href {https://doi.org/10.1103/PhysRevApplied.17.L061005} {\bibfield
  {journal} {\bibinfo  {journal} {Phys. Rev. Applied}\ }\textbf {\bibinfo
  {volume} {17}},\ \bibinfo {pages} {L061005} (\bibinfo {year}
  {2022})}\BibitemShut {NoStop}%
\bibitem [{\citenamefont {Dicke}(1946)}]{doi:10.1063/1.1770483}%
  \BibitemOpen
  \bibfield  {author} {\bibinfo {author} {\bibfnamefont {R.~H.}\ \bibnamefont
  {Dicke}},\ }\bibfield  {title} {\bibinfo {title} {The measurement of thermal
  radiation at microwave frequencies},\ }\href
  {https://doi.org/10.1063/1.1770483} {\bibfield  {journal} {\bibinfo
  {journal} {Review of Scientific Instruments}\ }\textbf {\bibinfo {volume}
  {17}},\ \bibinfo {pages} {268} (\bibinfo {year} {1946})},\ \Eprint
  {https://arxiv.org/abs/https://doi.org/10.1063/1.1770483}
  {https://doi.org/10.1063/1.1770483} \BibitemShut {NoStop}%
\bibitem [{\citenamefont {Peng}\ \emph {et~al.}(2000)\citenamefont {Peng} \emph
  {et~al.}}]{Peng:2000hd}%
  \BibitemOpen
  \bibfield  {author} {\bibinfo {author} {\bibfnamefont {H.}~\bibnamefont
  {Peng}} \emph {et~al.},\ }\bibfield  {title} {\bibinfo {title} {{Cryogenic
  cavity detector for a large scale cold dark-matter axion search}},\ }\href
  {https://doi.org/10.1016/S0168-9002(99)00971-7} {\bibfield  {journal}
  {\bibinfo  {journal} {Nucl. Instrum. Meth. A}\ }\textbf {\bibinfo {volume}
  {444}},\ \bibinfo {pages} {569} (\bibinfo {year} {2000})}\BibitemShut
  {NoStop}%
\bibitem [{\citenamefont {{Al Kenany}}\ \emph {et~al.}(2017)\citenamefont {{Al
  Kenany}}, \citenamefont {Anil}, \citenamefont {Backes}, \citenamefont
  {Brubaker}, \citenamefont {Cahn}, \citenamefont {Carosi}, \citenamefont
  {Gurevich}, \citenamefont {Kindel}, \citenamefont {Lamoreaux}, \citenamefont
  {Lehnert}, \citenamefont {Lewis}, \citenamefont {Malnou}, \citenamefont
  {Palken}, \citenamefont {Rapidis}, \citenamefont {Root}, \citenamefont
  {Simanovskaia}, \citenamefont {Shokair}, \citenamefont {Urdinaran},
  \citenamefont {{van Bibber}},\ and\ \citenamefont {Zhong}}]{ALKENANY201711}%
  \BibitemOpen
  \bibfield  {author} {\bibinfo {author} {\bibfnamefont {S.}~\bibnamefont {{Al
  Kenany}}}, \bibinfo {author} {\bibfnamefont {M.}~\bibnamefont {Anil}},
  \bibinfo {author} {\bibfnamefont {K.}~\bibnamefont {Backes}}, \bibinfo
  {author} {\bibfnamefont {B.}~\bibnamefont {Brubaker}}, \bibinfo {author}
  {\bibfnamefont {S.}~\bibnamefont {Cahn}}, \bibinfo {author} {\bibfnamefont
  {G.}~\bibnamefont {Carosi}}, \bibinfo {author} {\bibfnamefont
  {Y.}~\bibnamefont {Gurevich}}, \bibinfo {author} {\bibfnamefont
  {W.}~\bibnamefont {Kindel}}, \bibinfo {author} {\bibfnamefont
  {S.}~\bibnamefont {Lamoreaux}}, \bibinfo {author} {\bibfnamefont
  {K.}~\bibnamefont {Lehnert}}, \bibinfo {author} {\bibfnamefont
  {S.}~\bibnamefont {Lewis}}, \bibinfo {author} {\bibfnamefont
  {M.}~\bibnamefont {Malnou}}, \bibinfo {author} {\bibfnamefont
  {D.}~\bibnamefont {Palken}}, \bibinfo {author} {\bibfnamefont
  {N.}~\bibnamefont {Rapidis}}, \bibinfo {author} {\bibfnamefont
  {J.}~\bibnamefont {Root}}, \bibinfo {author} {\bibfnamefont {M.}~\bibnamefont
  {Simanovskaia}}, \bibinfo {author} {\bibfnamefont {T.}~\bibnamefont
  {Shokair}}, \bibinfo {author} {\bibfnamefont {I.}~\bibnamefont {Urdinaran}},
  \bibinfo {author} {\bibfnamefont {K.}~\bibnamefont {{van Bibber}}},\ and\
  \bibinfo {author} {\bibfnamefont {L.}~\bibnamefont {Zhong}},\ }\bibfield
  {title} {\bibinfo {title} {Design and operational experience of a microwave
  cavity axion detector for the 20–100 $\mu$ev range},\ }\href
  {https://doi.org/https://doi.org/10.1016/j.nima.2017.02.012} {\bibfield
  {journal} {\bibinfo  {journal} {Nuclear Instruments and Methods in Physics
  Research Section A: Accelerators, Spectrometers, Detectors and Associated
  Equipment}\ }\textbf {\bibinfo {volume} {854}},\ \bibinfo {pages} {11}
  (\bibinfo {year} {2017})}\BibitemShut {NoStop}%
\bibitem [{\citenamefont {Cervantes}\ \emph
  {et~al.}(2022{\natexlab{b}})\citenamefont {Cervantes}, \citenamefont
  {Carosi}, \citenamefont {Hanretty}, \citenamefont {Kimes}, \citenamefont
  {LaRoque}, \citenamefont {Leum}, \citenamefont {Mohapatra}, \citenamefont
  {Oblath}, \citenamefont {Ottens}, \citenamefont {Park}, \citenamefont
  {Rybka}, \citenamefont {Sinnis},\ and\ \citenamefont {Yang}}]{OrpheusPRL}%
  \BibitemOpen
  \bibfield  {author} {\bibinfo {author} {\bibfnamefont {R.}~\bibnamefont
  {Cervantes}}, \bibinfo {author} {\bibfnamefont {G.}~\bibnamefont {Carosi}},
  \bibinfo {author} {\bibfnamefont {C.}~\bibnamefont {Hanretty}}, \bibinfo
  {author} {\bibfnamefont {S.}~\bibnamefont {Kimes}}, \bibinfo {author}
  {\bibfnamefont {B.~H.}\ \bibnamefont {LaRoque}}, \bibinfo {author}
  {\bibfnamefont {G.}~\bibnamefont {Leum}}, \bibinfo {author} {\bibfnamefont
  {P.}~\bibnamefont {Mohapatra}}, \bibinfo {author} {\bibfnamefont {N.~S.}\
  \bibnamefont {Oblath}}, \bibinfo {author} {\bibfnamefont {R.}~\bibnamefont
  {Ottens}}, \bibinfo {author} {\bibfnamefont {Y.}~\bibnamefont {Park}},
  \bibinfo {author} {\bibfnamefont {G.}~\bibnamefont {Rybka}}, \bibinfo
  {author} {\bibfnamefont {J.}~\bibnamefont {Sinnis}},\ and\ \bibinfo {author}
  {\bibfnamefont {J.}~\bibnamefont {Yang}},\ }\href
  {https://doi.org/10.48550/ARXIV.2204.03818} {\bibinfo {title} {Search for 70
  $\mu$ev dark photon dark matter with a dielectrically-loaded multi-wavelength
  microwave cavity}} (\bibinfo {year} {2022}{\natexlab{b}}),\ \Eprint
  {https://arxiv.org/abs/2204.03818} {arXiv:2204.03818} \BibitemShut {NoStop}%
\bibitem [{Note1()}]{Note1}%
  \BibitemOpen
  \bibinfo {note} {Ref.~\cite {Kim_2020} also addresses the scan rate for
  ultrahigh Q haloscopes, but our treatment differs in a few major ways. For
  reasons explained in the text, we set the tuning step to $\Delta f \sim
  f_0/Q_{\protect \text {DM}}$ instead of $\Delta f \sim f_0/Q_{\protect \text
  {L}}$ and $b \sim f_0/Q_{\protect \text {L}}$ instead of $b \sim
  f_0/Q_{\protect \text {DM}}$. Finally, the derived $T_n$ in Ref.~\cite
  {Kim_2020} does not appear to apply systems ubiquitous to microwave haloscope
  experiments that implement circulators between the cavity and first-stage
  amplifier. For such a system, $T_n$ is independent of $\beta $~\cite
  {ALKENANY201711}. This independence is recognized in Fig.~5 but is not
  reflected in their equations.}\BibitemShut {Stop}%
\bibitem [{\citenamefont {Bartram}\ \emph
  {et~al.}(2021{\natexlab{a}})\citenamefont {Bartram}, \citenamefont {Braine},
  \citenamefont {Burns}, \citenamefont {Cervantes}, \citenamefont {Crisosto},
  \citenamefont {Du}, \citenamefont {Korandla}, \citenamefont {Leum},
  \citenamefont {Mohapatra}, \citenamefont {Nitta}, \citenamefont {Rosenberg},
  \citenamefont {Rybka}, \citenamefont {Yang}, \citenamefont {Clarke},
  \citenamefont {Siddiqi}, \citenamefont {Agrawal}, \citenamefont {Dixit},
  \citenamefont {Awida}, \citenamefont {Chou}, \citenamefont {Hollister},
  \citenamefont {Knirck}, \citenamefont {Sonnenschein}, \citenamefont {Wester},
  \citenamefont {Gleason}, \citenamefont {Hipp}, \citenamefont {Jois},
  \citenamefont {Sikivie}, \citenamefont {Sullivan}, \citenamefont {Tanner},
  \citenamefont {Lentz}, \citenamefont {Khatiwada}, \citenamefont {Carosi},
  \citenamefont {Robertson}, \citenamefont {Woollett}, \citenamefont {Duffy},
  \citenamefont {Boutan}, \citenamefont {Jones}, \citenamefont {LaRoque},
  \citenamefont {Oblath}, \citenamefont {Taubman}, \citenamefont {Daw},
  \citenamefont {Perry}, \citenamefont {Buckley}, \citenamefont {Gaikwad},
  \citenamefont {Hoffman}, \citenamefont {Murch}, \citenamefont {Goryachev},
  \citenamefont {McAllister}, \citenamefont {Quiskamp}, \citenamefont
  {Thomson},\ and\ \citenamefont {Tobar}}]{PhysRevLett.127.261803}%
  \BibitemOpen
  \bibfield  {author} {\bibinfo {author} {\bibfnamefont {C.}~\bibnamefont
  {Bartram}}, \bibinfo {author} {\bibfnamefont {T.}~\bibnamefont {Braine}},
  \bibinfo {author} {\bibfnamefont {E.}~\bibnamefont {Burns}}, \bibinfo
  {author} {\bibfnamefont {R.}~\bibnamefont {Cervantes}}, \bibinfo {author}
  {\bibfnamefont {N.}~\bibnamefont {Crisosto}}, \bibinfo {author}
  {\bibfnamefont {N.}~\bibnamefont {Du}}, \bibinfo {author} {\bibfnamefont
  {H.}~\bibnamefont {Korandla}}, \bibinfo {author} {\bibfnamefont
  {G.}~\bibnamefont {Leum}}, \bibinfo {author} {\bibfnamefont {P.}~\bibnamefont
  {Mohapatra}}, \bibinfo {author} {\bibfnamefont {T.}~\bibnamefont {Nitta}},
  \bibinfo {author} {\bibfnamefont {L.~J.}\ \bibnamefont {Rosenberg}}, \bibinfo
  {author} {\bibfnamefont {G.}~\bibnamefont {Rybka}}, \bibinfo {author}
  {\bibfnamefont {J.}~\bibnamefont {Yang}}, \bibinfo {author} {\bibfnamefont
  {J.}~\bibnamefont {Clarke}}, \bibinfo {author} {\bibfnamefont
  {I.}~\bibnamefont {Siddiqi}}, \bibinfo {author} {\bibfnamefont
  {A.}~\bibnamefont {Agrawal}}, \bibinfo {author} {\bibfnamefont {A.~V.}\
  \bibnamefont {Dixit}}, \bibinfo {author} {\bibfnamefont {M.~H.}\ \bibnamefont
  {Awida}}, \bibinfo {author} {\bibfnamefont {A.~S.}\ \bibnamefont {Chou}},
  \bibinfo {author} {\bibfnamefont {M.}~\bibnamefont {Hollister}}, \bibinfo
  {author} {\bibfnamefont {S.}~\bibnamefont {Knirck}}, \bibinfo {author}
  {\bibfnamefont {A.}~\bibnamefont {Sonnenschein}}, \bibinfo {author}
  {\bibfnamefont {W.}~\bibnamefont {Wester}}, \bibinfo {author} {\bibfnamefont
  {J.~R.}\ \bibnamefont {Gleason}}, \bibinfo {author} {\bibfnamefont {A.~T.}\
  \bibnamefont {Hipp}}, \bibinfo {author} {\bibfnamefont {S.}~\bibnamefont
  {Jois}}, \bibinfo {author} {\bibfnamefont {P.}~\bibnamefont {Sikivie}},
  \bibinfo {author} {\bibfnamefont {N.~S.}\ \bibnamefont {Sullivan}}, \bibinfo
  {author} {\bibfnamefont {D.~B.}\ \bibnamefont {Tanner}}, \bibinfo {author}
  {\bibfnamefont {E.}~\bibnamefont {Lentz}}, \bibinfo {author} {\bibfnamefont
  {R.}~\bibnamefont {Khatiwada}}, \bibinfo {author} {\bibfnamefont
  {G.}~\bibnamefont {Carosi}}, \bibinfo {author} {\bibfnamefont
  {N.}~\bibnamefont {Robertson}}, \bibinfo {author} {\bibfnamefont
  {N.}~\bibnamefont {Woollett}}, \bibinfo {author} {\bibfnamefont {L.~D.}\
  \bibnamefont {Duffy}}, \bibinfo {author} {\bibfnamefont {C.}~\bibnamefont
  {Boutan}}, \bibinfo {author} {\bibfnamefont {M.}~\bibnamefont {Jones}},
  \bibinfo {author} {\bibfnamefont {B.~H.}\ \bibnamefont {LaRoque}}, \bibinfo
  {author} {\bibfnamefont {N.~S.}\ \bibnamefont {Oblath}}, \bibinfo {author}
  {\bibfnamefont {M.~S.}\ \bibnamefont {Taubman}}, \bibinfo {author}
  {\bibfnamefont {E.~J.}\ \bibnamefont {Daw}}, \bibinfo {author} {\bibfnamefont
  {M.~G.}\ \bibnamefont {Perry}}, \bibinfo {author} {\bibfnamefont {J.~H.}\
  \bibnamefont {Buckley}}, \bibinfo {author} {\bibfnamefont {C.}~\bibnamefont
  {Gaikwad}}, \bibinfo {author} {\bibfnamefont {J.}~\bibnamefont {Hoffman}},
  \bibinfo {author} {\bibfnamefont {K.~W.}\ \bibnamefont {Murch}}, \bibinfo
  {author} {\bibfnamefont {M.}~\bibnamefont {Goryachev}}, \bibinfo {author}
  {\bibfnamefont {B.~T.}\ \bibnamefont {McAllister}}, \bibinfo {author}
  {\bibfnamefont {A.}~\bibnamefont {Quiskamp}}, \bibinfo {author}
  {\bibfnamefont {C.}~\bibnamefont {Thomson}},\ and\ \bibinfo {author}
  {\bibfnamefont {M.~E.}\ \bibnamefont {Tobar}} (\bibinfo {collaboration} {ADMX
  Collaboration}),\ }\bibfield  {title} {\bibinfo {title} {Search for invisible
  axion dark matter in the $3.3--4.2\text{ }\text{
  }\ensuremath{\mu}\mathrm{eV}$ mass range},\ }\href
  {https://doi.org/10.1103/PhysRevLett.127.261803} {\bibfield  {journal}
  {\bibinfo  {journal} {Phys. Rev. Lett.}\ }\textbf {\bibinfo {volume} {127}},\
  \bibinfo {pages} {261803} (\bibinfo {year} {2021}{\natexlab{a}})}\BibitemShut
  {NoStop}%
\bibitem [{\citenamefont {Aune}\ \emph {et~al.}(2000)\citenamefont {Aune} \emph
  {et~al.}}]{Aune:2000gb}%
  \BibitemOpen
  \bibfield  {author} {\bibinfo {author} {\bibfnamefont {B.}~\bibnamefont
  {Aune}} \emph {et~al.},\ }\bibfield  {title} {\bibinfo {title} {{The
  superconducting TESLA cavities}},\ }\href
  {https://doi.org/10.1103/PhysRevSTAB.3.092001} {\bibfield  {journal}
  {\bibinfo  {journal} {Phys. Rev. ST Accel. Beams}\ }\textbf {\bibinfo
  {volume} {3}},\ \bibinfo {pages} {092001} (\bibinfo {year} {2000})},\ \Eprint
  {https://arxiv.org/abs/physics/0003011} {arXiv:physics/0003011} \BibitemShut
  {NoStop}%
\bibitem [{\citenamefont {Romanenko}\ and\ \citenamefont
  {Schuster}(2017)}]{PhysRevLett.119.264801}%
  \BibitemOpen
  \bibfield  {author} {\bibinfo {author} {\bibfnamefont {A.}~\bibnamefont
  {Romanenko}}\ and\ \bibinfo {author} {\bibfnamefont {D.~I.}\ \bibnamefont
  {Schuster}},\ }\bibfield  {title} {\bibinfo {title} {Understanding quality
  factor degradation in superconducting niobium cavities at low microwave field
  amplitudes},\ }\href {https://doi.org/10.1103/PhysRevLett.119.264801}
  {\bibfield  {journal} {\bibinfo  {journal} {Phys. Rev. Lett.}\ }\textbf
  {\bibinfo {volume} {119}},\ \bibinfo {pages} {264801} (\bibinfo {year}
  {2017})}\BibitemShut {NoStop}%
\bibitem [{\citenamefont {Posen}\ \emph {et~al.}(2020)\citenamefont {Posen},
  \citenamefont {Romanenko}, \citenamefont {Grassellino}, \citenamefont
  {Melnychuk},\ and\ \citenamefont {Sergatskov}}]{PhysRevApplied.13.014024}%
  \BibitemOpen
  \bibfield  {author} {\bibinfo {author} {\bibfnamefont {S.}~\bibnamefont
  {Posen}}, \bibinfo {author} {\bibfnamefont {A.}~\bibnamefont {Romanenko}},
  \bibinfo {author} {\bibfnamefont {A.}~\bibnamefont {Grassellino}}, \bibinfo
  {author} {\bibfnamefont {O.}~\bibnamefont {Melnychuk}},\ and\ \bibinfo
  {author} {\bibfnamefont {D.}~\bibnamefont {Sergatskov}},\ }\bibfield  {title}
  {\bibinfo {title} {Ultralow surface resistance via vacuum heat treatment of
  superconducting radio-frequency cavities},\ }\href
  {https://doi.org/10.1103/PhysRevApplied.13.014024} {\bibfield  {journal}
  {\bibinfo  {journal} {Phys. Rev. Applied}\ }\textbf {\bibinfo {volume}
  {13}},\ \bibinfo {pages} {014024} (\bibinfo {year} {2020})}\BibitemShut
  {NoStop}%
\bibitem [{\citenamefont {Simbierowicz}\ \emph {et~al.}(2021)\citenamefont
  {Simbierowicz}, \citenamefont {Vesterinen}, \citenamefont {Milem},
  \citenamefont {Lintunen}, \citenamefont {Oksanen}, \citenamefont {Roschier},
  \citenamefont {Grönberg}, \citenamefont {Hassel}, \citenamefont
  {Gunnarsson},\ and\ \citenamefont {Lake}}]{10.1063/5.0028951}%
  \BibitemOpen
  \bibfield  {author} {\bibinfo {author} {\bibfnamefont {S.}~\bibnamefont
  {Simbierowicz}}, \bibinfo {author} {\bibfnamefont {V.}~\bibnamefont
  {Vesterinen}}, \bibinfo {author} {\bibfnamefont {J.}~\bibnamefont {Milem}},
  \bibinfo {author} {\bibfnamefont {A.}~\bibnamefont {Lintunen}}, \bibinfo
  {author} {\bibfnamefont {M.}~\bibnamefont {Oksanen}}, \bibinfo {author}
  {\bibfnamefont {L.}~\bibnamefont {Roschier}}, \bibinfo {author}
  {\bibfnamefont {L.}~\bibnamefont {Grönberg}}, \bibinfo {author}
  {\bibfnamefont {J.}~\bibnamefont {Hassel}}, \bibinfo {author} {\bibfnamefont
  {D.}~\bibnamefont {Gunnarsson}},\ and\ \bibinfo {author} {\bibfnamefont
  {R.~E.}\ \bibnamefont {Lake}},\ }\bibfield  {title} {\bibinfo {title}
  {{Characterizing cryogenic amplifiers with a matched temperature-variable
  noise source}},\ }\href {https://doi.org/10.1063/5.0028951} {\bibfield
  {journal} {\bibinfo  {journal} {Review of Scientific Instruments}\ }\textbf
  {\bibinfo {volume} {92}},\ \bibinfo {pages} {034708} (\bibinfo {year}
  {2021})},\ \Eprint
  {https://arxiv.org/abs/https://pubs.aip.org/aip/rsi/article-pdf/doi/10.1063/5.0028951/13865090/034708\_1\_online.pdf}
  {https://pubs.aip.org/aip/rsi/article-pdf/doi/10.1063/5.0028951/13865090/034708\_1\_online.pdf}
  \BibitemShut {NoStop}%
\bibitem [{\citenamefont {Pozar}(2012)}]{pozar}%
  \BibitemOpen
  \bibfield  {author} {\bibinfo {author} {\bibfnamefont {D.~M.}\ \bibnamefont
  {Pozar}},\ }\href@noop {} {\emph {\bibinfo {title} {Microwave Engineering}}}\
  (\bibinfo  {publisher} {John Wiley \& Sons, Inc},\ \bibinfo {address}
  {Hoboken, New Jersey},\ \bibinfo {year} {2012})\BibitemShut {NoStop}%
\bibitem [{\citenamefont {Fong}\ \emph {et~al.}(2011)\citenamefont {Fong},
  \citenamefont {Laverty}, \citenamefont {Chojnacki}, \citenamefont {Wang},\
  and\ \citenamefont {Hoffstaetter}}]{Fong2011SELFEO}%
  \BibitemOpen
  \bibfield  {author} {\bibinfo {author} {\bibfnamefont {K.}~\bibnamefont
  {Fong}}, \bibinfo {author} {\bibfnamefont {M.}~\bibnamefont {Laverty}},
  \bibinfo {author} {\bibfnamefont {E.}~\bibnamefont {Chojnacki}}, \bibinfo
  {author} {\bibfnamefont {S.~P.}\ \bibnamefont {Wang}},\ and\ \bibinfo
  {author} {\bibfnamefont {G.~H.}\ \bibnamefont {Hoffstaetter}},\ }\bibfield
  {title} {\bibinfo {title} {Self excited operation for a 1.3 ghz 5-cell
  superconducting cavity}\ }(\bibinfo {year} {2011})\BibitemShut {NoStop}%
\bibitem [{\citenamefont {Delayen}(1978)}]{delayen1978phase}%
  \BibitemOpen
  \bibfield  {author} {\bibinfo {author} {\bibfnamefont {J.~R.}\ \bibnamefont
  {Delayen}},\ }\href@noop {} {\emph {\bibinfo {title} {Phase and amplitude
  stabilization of superconducting resonators.}}}\ (\bibinfo  {publisher}
  {California Institute of Technology},\ \bibinfo {year} {1978})\BibitemShut
  {NoStop}%
\bibitem [{\citenamefont {Padamsee}\ \emph {et~al.}(2008)\citenamefont
  {Padamsee}, \citenamefont {Knobloch},\ and\ \citenamefont {Hays}}]{padamsee}%
  \BibitemOpen
  \bibfield  {author} {\bibinfo {author} {\bibfnamefont {H.}~\bibnamefont
  {Padamsee}}, \bibinfo {author} {\bibfnamefont {J.}~\bibnamefont {Knobloch}},\
  and\ \bibinfo {author} {\bibfnamefont {T.}~\bibnamefont {Hays}},\ }\href@noop
  {} {\emph {\bibinfo {title} {RF Superconductivity for Accelerators, 2nd
  Edition}}}\ (\bibinfo  {publisher} {Wiley-VCH},\ \bibinfo {year}
  {2008})\BibitemShut {NoStop}%
\bibitem [{\citenamefont {Melnychuk}\ \emph {et~al.}(2014)\citenamefont
  {Melnychuk}, \citenamefont {Grassellino},\ and\ \citenamefont
  {Romanenko}}]{doi:10.1063/1.4903868}%
  \BibitemOpen
  \bibfield  {author} {\bibinfo {author} {\bibfnamefont {O.}~\bibnamefont
  {Melnychuk}}, \bibinfo {author} {\bibfnamefont {A.}~\bibnamefont
  {Grassellino}},\ and\ \bibinfo {author} {\bibfnamefont {A.}~\bibnamefont
  {Romanenko}},\ }\bibfield  {title} {\bibinfo {title} {Error analysis for
  intrinsic quality factor measurement in superconducting radio frequency
  resonators},\ }\href {https://doi.org/10.1063/1.4903868} {\bibfield
  {journal} {\bibinfo  {journal} {Review of Scientific Instruments}\ }\textbf
  {\bibinfo {volume} {85}},\ \bibinfo {pages} {124705} (\bibinfo {year}
  {2014})},\ \Eprint {https://arxiv.org/abs/https://doi.org/10.1063/1.4903868}
  {https://doi.org/10.1063/1.4903868} \BibitemShut {NoStop}%
\bibitem [{\citenamefont {Brubaker}\ \emph {et~al.}(2017)\citenamefont
  {Brubaker}, \citenamefont {Zhong}, \citenamefont {Lamoreaux}, \citenamefont
  {Lehnert},\ and\ \citenamefont {van Bibber}}]{PhysRevD.96.123008}%
  \BibitemOpen
  \bibfield  {author} {\bibinfo {author} {\bibfnamefont {B.~M.}\ \bibnamefont
  {Brubaker}}, \bibinfo {author} {\bibfnamefont {L.}~\bibnamefont {Zhong}},
  \bibinfo {author} {\bibfnamefont {S.~K.}\ \bibnamefont {Lamoreaux}}, \bibinfo
  {author} {\bibfnamefont {K.~W.}\ \bibnamefont {Lehnert}},\ and\ \bibinfo
  {author} {\bibfnamefont {K.~A.}\ \bibnamefont {van Bibber}},\ }\bibfield
  {title} {\bibinfo {title} {Haystac axion search analysis procedure},\ }\href
  {https://doi.org/10.1103/PhysRevD.96.123008} {\bibfield  {journal} {\bibinfo
  {journal} {Phys. Rev. D}\ }\textbf {\bibinfo {volume} {96}},\ \bibinfo
  {pages} {123008} (\bibinfo {year} {2017})}\BibitemShut {NoStop}%
\bibitem [{\citenamefont {Bartram}\ \emph
  {et~al.}(2021{\natexlab{b}})\citenamefont {Bartram}, \citenamefont {Braine},
  \citenamefont {Cervantes}, \citenamefont {Crisosto}, \citenamefont {Du},
  \citenamefont {Leum}, \citenamefont {Rosenberg}, \citenamefont {Rybka},
  \citenamefont {Yang}, \citenamefont {Bowring}, \citenamefont {Chou},
  \citenamefont {Khatiwada}, \citenamefont {Sonnenschein}, \citenamefont
  {Wester}, \citenamefont {Carosi}, \citenamefont {Woollett}, \citenamefont
  {Duffy}, \citenamefont {Goryachev}, \citenamefont {McAllister}, \citenamefont
  {Tobar}, \citenamefont {Boutan}, \citenamefont {Jones}, \citenamefont
  {LaRoque}, \citenamefont {Oblath}, \citenamefont {Taubman}, \citenamefont
  {Clarke}, \citenamefont {Dove}, \citenamefont {Eddins}, \citenamefont
  {O'Kelley}, \citenamefont {Nawaz}, \citenamefont {Siddiqi}, \citenamefont
  {Stevenson}, \citenamefont {Agrawal}, \citenamefont {Dixit}, \citenamefont
  {Gleason}, \citenamefont {Jois}, \citenamefont {Sikivie}, \citenamefont
  {Solomon}, \citenamefont {Sullivan}, \citenamefont {Tanner}, \citenamefont
  {Lentz}, \citenamefont {Daw}, \citenamefont {Perry}, \citenamefont {Buckley},
  \citenamefont {Harrington}, \citenamefont {Henriksen},\ and\ \citenamefont
  {Murch}}]{PhysRevD.103.032002}%
  \BibitemOpen
  \bibfield  {author} {\bibinfo {author} {\bibfnamefont {C.}~\bibnamefont
  {Bartram}}, \bibinfo {author} {\bibfnamefont {T.}~\bibnamefont {Braine}},
  \bibinfo {author} {\bibfnamefont {R.}~\bibnamefont {Cervantes}}, \bibinfo
  {author} {\bibfnamefont {N.}~\bibnamefont {Crisosto}}, \bibinfo {author}
  {\bibfnamefont {N.}~\bibnamefont {Du}}, \bibinfo {author} {\bibfnamefont
  {G.}~\bibnamefont {Leum}}, \bibinfo {author} {\bibfnamefont {L.~J.}\
  \bibnamefont {Rosenberg}}, \bibinfo {author} {\bibfnamefont {G.}~\bibnamefont
  {Rybka}}, \bibinfo {author} {\bibfnamefont {J.}~\bibnamefont {Yang}},
  \bibinfo {author} {\bibfnamefont {D.}~\bibnamefont {Bowring}}, \bibinfo
  {author} {\bibfnamefont {A.~S.}\ \bibnamefont {Chou}}, \bibinfo {author}
  {\bibfnamefont {R.}~\bibnamefont {Khatiwada}}, \bibinfo {author}
  {\bibfnamefont {A.}~\bibnamefont {Sonnenschein}}, \bibinfo {author}
  {\bibfnamefont {W.}~\bibnamefont {Wester}}, \bibinfo {author} {\bibfnamefont
  {G.}~\bibnamefont {Carosi}}, \bibinfo {author} {\bibfnamefont
  {N.}~\bibnamefont {Woollett}}, \bibinfo {author} {\bibfnamefont {L.~D.}\
  \bibnamefont {Duffy}}, \bibinfo {author} {\bibfnamefont {M.}~\bibnamefont
  {Goryachev}}, \bibinfo {author} {\bibfnamefont {B.}~\bibnamefont
  {McAllister}}, \bibinfo {author} {\bibfnamefont {M.~E.}\ \bibnamefont
  {Tobar}}, \bibinfo {author} {\bibfnamefont {C.}~\bibnamefont {Boutan}},
  \bibinfo {author} {\bibfnamefont {M.}~\bibnamefont {Jones}}, \bibinfo
  {author} {\bibfnamefont {B.~H.}\ \bibnamefont {LaRoque}}, \bibinfo {author}
  {\bibfnamefont {N.~S.}\ \bibnamefont {Oblath}}, \bibinfo {author}
  {\bibfnamefont {M.~S.}\ \bibnamefont {Taubman}}, \bibinfo {author}
  {\bibfnamefont {J.}~\bibnamefont {Clarke}}, \bibinfo {author} {\bibfnamefont
  {A.}~\bibnamefont {Dove}}, \bibinfo {author} {\bibfnamefont {A.}~\bibnamefont
  {Eddins}}, \bibinfo {author} {\bibfnamefont {S.~R.}\ \bibnamefont
  {O'Kelley}}, \bibinfo {author} {\bibfnamefont {S.}~\bibnamefont {Nawaz}},
  \bibinfo {author} {\bibfnamefont {I.}~\bibnamefont {Siddiqi}}, \bibinfo
  {author} {\bibfnamefont {N.}~\bibnamefont {Stevenson}}, \bibinfo {author}
  {\bibfnamefont {A.}~\bibnamefont {Agrawal}}, \bibinfo {author} {\bibfnamefont
  {A.~V.}\ \bibnamefont {Dixit}}, \bibinfo {author} {\bibfnamefont {J.~R.}\
  \bibnamefont {Gleason}}, \bibinfo {author} {\bibfnamefont {S.}~\bibnamefont
  {Jois}}, \bibinfo {author} {\bibfnamefont {P.}~\bibnamefont {Sikivie}},
  \bibinfo {author} {\bibfnamefont {J.~A.}\ \bibnamefont {Solomon}}, \bibinfo
  {author} {\bibfnamefont {N.~S.}\ \bibnamefont {Sullivan}}, \bibinfo {author}
  {\bibfnamefont {D.~B.}\ \bibnamefont {Tanner}}, \bibinfo {author}
  {\bibfnamefont {E.}~\bibnamefont {Lentz}}, \bibinfo {author} {\bibfnamefont
  {E.~J.}\ \bibnamefont {Daw}}, \bibinfo {author} {\bibfnamefont {M.~G.}\
  \bibnamefont {Perry}}, \bibinfo {author} {\bibfnamefont {J.~H.}\ \bibnamefont
  {Buckley}}, \bibinfo {author} {\bibfnamefont {P.~M.}\ \bibnamefont
  {Harrington}}, \bibinfo {author} {\bibfnamefont {E.~A.}\ \bibnamefont
  {Henriksen}},\ and\ \bibinfo {author} {\bibfnamefont {K.~W.}\ \bibnamefont
  {Murch}} (\bibinfo {collaboration} {ADMX Collaboration}),\ }\bibfield
  {title} {\bibinfo {title} {Axion dark matter experiment: Run 1b analysis
  details},\ }\href {https://doi.org/10.1103/PhysRevD.103.032002} {\bibfield
  {journal} {\bibinfo  {journal} {Phys. Rev. D}\ }\textbf {\bibinfo {volume}
  {103}},\ \bibinfo {pages} {032002} (\bibinfo {year}
  {2021}{\natexlab{b}})}\BibitemShut {NoStop}%
\bibitem [{\citenamefont {Asztalos}\ \emph {et~al.}(2001)\citenamefont
  {Asztalos}, \citenamefont {Daw}, \citenamefont {Peng}, \citenamefont
  {Rosenberg}, \citenamefont {Hagmann}, \citenamefont {Kinion}, \citenamefont
  {Stoeffl}, \citenamefont {van Bibber}, \citenamefont {Sikivie}, \citenamefont
  {Sullivan}, \citenamefont {Tanner}, \citenamefont {Nezrick}, \citenamefont
  {Turner}, \citenamefont {Moltz}, \citenamefont {Powell}, \citenamefont
  {Andr\'e}, \citenamefont {Clarke}, \citenamefont {M\"uck},\ and\
  \citenamefont {Bradley}}]{PhysRevD.64.092003}%
  \BibitemOpen
  \bibfield  {author} {\bibinfo {author} {\bibfnamefont {S.}~\bibnamefont
  {Asztalos}}, \bibinfo {author} {\bibfnamefont {E.}~\bibnamefont {Daw}},
  \bibinfo {author} {\bibfnamefont {H.}~\bibnamefont {Peng}}, \bibinfo {author}
  {\bibfnamefont {L.~J.}\ \bibnamefont {Rosenberg}}, \bibinfo {author}
  {\bibfnamefont {C.}~\bibnamefont {Hagmann}}, \bibinfo {author} {\bibfnamefont
  {D.}~\bibnamefont {Kinion}}, \bibinfo {author} {\bibfnamefont
  {W.}~\bibnamefont {Stoeffl}}, \bibinfo {author} {\bibfnamefont
  {K.}~\bibnamefont {van Bibber}}, \bibinfo {author} {\bibfnamefont
  {P.}~\bibnamefont {Sikivie}}, \bibinfo {author} {\bibfnamefont {N.~S.}\
  \bibnamefont {Sullivan}}, \bibinfo {author} {\bibfnamefont {D.~B.}\
  \bibnamefont {Tanner}}, \bibinfo {author} {\bibfnamefont {F.}~\bibnamefont
  {Nezrick}}, \bibinfo {author} {\bibfnamefont {M.~S.}\ \bibnamefont {Turner}},
  \bibinfo {author} {\bibfnamefont {D.~M.}\ \bibnamefont {Moltz}}, \bibinfo
  {author} {\bibfnamefont {J.}~\bibnamefont {Powell}}, \bibinfo {author}
  {\bibfnamefont {M.-O.}\ \bibnamefont {Andr\'e}}, \bibinfo {author}
  {\bibfnamefont {J.}~\bibnamefont {Clarke}}, \bibinfo {author} {\bibfnamefont
  {M.}~\bibnamefont {M\"uck}},\ and\ \bibinfo {author} {\bibfnamefont {R.~F.}\
  \bibnamefont {Bradley}},\ }\bibfield  {title} {\bibinfo {title} {Large-scale
  microwave cavity search for dark-matter axions},\ }\href
  {https://doi.org/10.1103/PhysRevD.64.092003} {\bibfield  {journal} {\bibinfo
  {journal} {Phys. Rev. D}\ }\textbf {\bibinfo {volume} {64}},\ \bibinfo
  {pages} {092003} (\bibinfo {year} {2001})}\BibitemShut {NoStop}%
\bibitem [{\citenamefont {Dixit}\ \emph {et~al.}(2021)\citenamefont {Dixit},
  \citenamefont {Chakram}, \citenamefont {He}, \citenamefont {Agrawal},
  \citenamefont {Naik}, \citenamefont {Schuster},\ and\ \citenamefont
  {Chou}}]{PhysRevLett.126.141302}%
  \BibitemOpen
  \bibfield  {author} {\bibinfo {author} {\bibfnamefont {A.~V.}\ \bibnamefont
  {Dixit}}, \bibinfo {author} {\bibfnamefont {S.}~\bibnamefont {Chakram}},
  \bibinfo {author} {\bibfnamefont {K.}~\bibnamefont {He}}, \bibinfo {author}
  {\bibfnamefont {A.}~\bibnamefont {Agrawal}}, \bibinfo {author} {\bibfnamefont
  {R.~K.}\ \bibnamefont {Naik}}, \bibinfo {author} {\bibfnamefont {D.~I.}\
  \bibnamefont {Schuster}},\ and\ \bibinfo {author} {\bibfnamefont
  {A.}~\bibnamefont {Chou}},\ }\bibfield  {title} {\bibinfo {title} {Searching
  for dark matter with a superconducting qubit},\ }\href
  {https://doi.org/10.1103/PhysRevLett.126.141302} {\bibfield  {journal}
  {\bibinfo  {journal} {Phys. Rev. Lett.}\ }\textbf {\bibinfo {volume} {126}},\
  \bibinfo {pages} {141302} (\bibinfo {year} {2021})}\BibitemShut {NoStop}%
\bibitem [{\citenamefont {O'Hare}(2020)}]{ciaran_o_hare_2020_3932430}%
  \BibitemOpen
  \bibfield  {author} {\bibinfo {author} {\bibfnamefont {C.}~\bibnamefont
  {O'Hare}},\ }\href {https://doi.org/10.5281/zenodo.3932430} {\bibinfo {title}
  {cajohare/axionlimits: Axionlimits}} (\bibinfo {year} {2020})\BibitemShut
  {NoStop}%
\bibitem [{\citenamefont {Peccei}\ and\ \citenamefont
  {Quinn}(1977)}]{PhysRevLett.38.1440}%
  \BibitemOpen
  \bibfield  {author} {\bibinfo {author} {\bibfnamefont {R.~D.}\ \bibnamefont
  {Peccei}}\ and\ \bibinfo {author} {\bibfnamefont {H.~R.}\ \bibnamefont
  {Quinn}},\ }\bibfield  {title} {\bibinfo {title} {Cp conservation in the
  presence of pseudoparticles},\ }\href
  {https://doi.org/10.1103/PhysRevLett.38.1440} {\bibfield  {journal} {\bibinfo
   {journal} {Phys. Rev. Lett.}\ }\textbf {\bibinfo {volume} {38}},\ \bibinfo
  {pages} {1440} (\bibinfo {year} {1977})}\BibitemShut {NoStop}%
\bibitem [{\citenamefont {Backes}\ \emph {et~al.}(2021)\citenamefont {Backes},
  \citenamefont {Palken}, \citenamefont {Kenany}, \citenamefont {Brubaker},
  \citenamefont {Cahn}, \citenamefont {Droster}, \citenamefont {Hilton},
  \citenamefont {Ghosh}, \citenamefont {Jackson}, \citenamefont {Lamoreaux},
  \citenamefont {Leder}, \citenamefont {Lehnert}, \citenamefont {Lewis},
  \citenamefont {Malnou}, \citenamefont {Maruyama}, \citenamefont {Rapidis},
  \citenamefont {Simanovskaia}, \citenamefont {Singh}, \citenamefont {Speller},
  \citenamefont {Urdinaran}, \citenamefont {Vale}, \citenamefont {van
  Assendelft}, \citenamefont {van Bibber},\ and\ \citenamefont
  {Wang}}]{Backes2021}%
  \BibitemOpen
  \bibfield  {author} {\bibinfo {author} {\bibfnamefont {K.~M.}\ \bibnamefont
  {Backes}}, \bibinfo {author} {\bibfnamefont {D.~A.}\ \bibnamefont {Palken}},
  \bibinfo {author} {\bibfnamefont {S.~A.}\ \bibnamefont {Kenany}}, \bibinfo
  {author} {\bibfnamefont {B.~M.}\ \bibnamefont {Brubaker}}, \bibinfo {author}
  {\bibfnamefont {S.~B.}\ \bibnamefont {Cahn}}, \bibinfo {author}
  {\bibfnamefont {A.}~\bibnamefont {Droster}}, \bibinfo {author} {\bibfnamefont
  {G.~C.}\ \bibnamefont {Hilton}}, \bibinfo {author} {\bibfnamefont
  {S.}~\bibnamefont {Ghosh}}, \bibinfo {author} {\bibfnamefont
  {H.}~\bibnamefont {Jackson}}, \bibinfo {author} {\bibfnamefont {S.~K.}\
  \bibnamefont {Lamoreaux}}, \bibinfo {author} {\bibfnamefont {A.~F.}\
  \bibnamefont {Leder}}, \bibinfo {author} {\bibfnamefont {K.~W.}\ \bibnamefont
  {Lehnert}}, \bibinfo {author} {\bibfnamefont {S.~M.}\ \bibnamefont {Lewis}},
  \bibinfo {author} {\bibfnamefont {M.}~\bibnamefont {Malnou}}, \bibinfo
  {author} {\bibfnamefont {R.~H.}\ \bibnamefont {Maruyama}}, \bibinfo {author}
  {\bibfnamefont {N.~M.}\ \bibnamefont {Rapidis}}, \bibinfo {author}
  {\bibfnamefont {M.}~\bibnamefont {Simanovskaia}}, \bibinfo {author}
  {\bibfnamefont {S.}~\bibnamefont {Singh}}, \bibinfo {author} {\bibfnamefont
  {D.~H.}\ \bibnamefont {Speller}}, \bibinfo {author} {\bibfnamefont
  {I.}~\bibnamefont {Urdinaran}}, \bibinfo {author} {\bibfnamefont {L.~R.}\
  \bibnamefont {Vale}}, \bibinfo {author} {\bibfnamefont {E.~C.}\ \bibnamefont
  {van Assendelft}}, \bibinfo {author} {\bibfnamefont {K.}~\bibnamefont {van
  Bibber}},\ and\ \bibinfo {author} {\bibfnamefont {H.}~\bibnamefont {Wang}},\
  }\bibfield  {title} {\bibinfo {title} {A quantum enhanced search for dark
  matter axions},\ }\href {https://doi.org/10.1038/s41586-021-03226-7}
  {\bibfield  {journal} {\bibinfo  {journal} {Nature}\ }\textbf {\bibinfo
  {volume} {590}},\ \bibinfo {pages} {238} (\bibinfo {year}
  {2021})}\BibitemShut {NoStop}%
\bibitem [{\citenamefont {Cervantes}\ \emph
  {et~al.}(2022{\natexlab{c}})\citenamefont {Cervantes}, \citenamefont
  {Carosi}, \citenamefont {Hanretty}, \citenamefont {Kimes}, \citenamefont
  {LaRoque}, \citenamefont {Leum}, \citenamefont {Mohapatra}, \citenamefont
  {Oblath}, \citenamefont {Ottens}, \citenamefont {Park}, \citenamefont
  {Rybka}, \citenamefont {Sinnis},\ and\ \citenamefont
  {Yang}}]{PhysRevLett.129.201301}%
  \BibitemOpen
  \bibfield  {author} {\bibinfo {author} {\bibfnamefont {R.}~\bibnamefont
  {Cervantes}}, \bibinfo {author} {\bibfnamefont {G.}~\bibnamefont {Carosi}},
  \bibinfo {author} {\bibfnamefont {C.}~\bibnamefont {Hanretty}}, \bibinfo
  {author} {\bibfnamefont {S.}~\bibnamefont {Kimes}}, \bibinfo {author}
  {\bibfnamefont {B.~H.}\ \bibnamefont {LaRoque}}, \bibinfo {author}
  {\bibfnamefont {G.}~\bibnamefont {Leum}}, \bibinfo {author} {\bibfnamefont
  {P.}~\bibnamefont {Mohapatra}}, \bibinfo {author} {\bibfnamefont {N.~S.}\
  \bibnamefont {Oblath}}, \bibinfo {author} {\bibfnamefont {R.}~\bibnamefont
  {Ottens}}, \bibinfo {author} {\bibfnamefont {Y.}~\bibnamefont {Park}},
  \bibinfo {author} {\bibfnamefont {G.}~\bibnamefont {Rybka}}, \bibinfo
  {author} {\bibfnamefont {J.}~\bibnamefont {Sinnis}},\ and\ \bibinfo {author}
  {\bibfnamefont {J.}~\bibnamefont {Yang}},\ }\bibfield  {title} {\bibinfo
  {title} {Search for $70\text{ }\text{ }\ensuremath{\mu}\mathrm{eV}$ dark
  photon dark matter with a dielectrically loaded multiwavelength microwave
  cavity},\ }\href {https://doi.org/10.1103/PhysRevLett.129.201301} {\bibfield
  {journal} {\bibinfo  {journal} {Phys. Rev. Lett.}\ }\textbf {\bibinfo
  {volume} {129}},\ \bibinfo {pages} {201301} (\bibinfo {year}
  {2022}{\natexlab{c}})}\BibitemShut {NoStop}%
\bibitem [{\citenamefont {Brun}\ \emph {et~al.}(2019)\citenamefont {Brun},
  \citenamefont {Caldwell}, \citenamefont {Chevalier}, \citenamefont {Dvali},
  \citenamefont {Freire}, \citenamefont {Garutti}, \citenamefont {Heyminck},
  \citenamefont {Jochum}, \citenamefont {Knirck}, \citenamefont {Kramer},
  \citenamefont {Krieger}, \citenamefont {Lasserre}, \citenamefont {Lee},
  \citenamefont {Li}, \citenamefont {Lindner}, \citenamefont {Majorovits},
  \citenamefont {Martens}, \citenamefont {Matysek}, \citenamefont {Millar},
  \citenamefont {Raffelt}, \citenamefont {Redondo}, \citenamefont {Reimann},
  \citenamefont {Ringwald}, \citenamefont {Saikawa}, \citenamefont {Schaffran},
  \citenamefont {Schmidt}, \citenamefont {Sch{\"u}tte-Engel}, \citenamefont
  {Steffen}, \citenamefont {Strandhagen}, \citenamefont {Wieching},\ and\
  \citenamefont {Collaboration}}]{Brun2019}%
  \BibitemOpen
  \bibfield  {author} {\bibinfo {author} {\bibfnamefont {P.}~\bibnamefont
  {Brun}}, \bibinfo {author} {\bibfnamefont {A.}~\bibnamefont {Caldwell}},
  \bibinfo {author} {\bibfnamefont {L.}~\bibnamefont {Chevalier}}, \bibinfo
  {author} {\bibfnamefont {G.}~\bibnamefont {Dvali}}, \bibinfo {author}
  {\bibfnamefont {P.}~\bibnamefont {Freire}}, \bibinfo {author} {\bibfnamefont
  {E.}~\bibnamefont {Garutti}}, \bibinfo {author} {\bibfnamefont
  {S.}~\bibnamefont {Heyminck}}, \bibinfo {author} {\bibfnamefont
  {J.}~\bibnamefont {Jochum}}, \bibinfo {author} {\bibfnamefont
  {S.}~\bibnamefont {Knirck}}, \bibinfo {author} {\bibfnamefont
  {M.}~\bibnamefont {Kramer}}, \bibinfo {author} {\bibfnamefont
  {C.}~\bibnamefont {Krieger}}, \bibinfo {author} {\bibfnamefont
  {T.}~\bibnamefont {Lasserre}}, \bibinfo {author} {\bibfnamefont
  {C.}~\bibnamefont {Lee}}, \bibinfo {author} {\bibfnamefont {X.}~\bibnamefont
  {Li}}, \bibinfo {author} {\bibfnamefont {A.}~\bibnamefont {Lindner}},
  \bibinfo {author} {\bibfnamefont {B.}~\bibnamefont {Majorovits}}, \bibinfo
  {author} {\bibfnamefont {S.}~\bibnamefont {Martens}}, \bibinfo {author}
  {\bibfnamefont {M.}~\bibnamefont {Matysek}}, \bibinfo {author} {\bibfnamefont
  {A.}~\bibnamefont {Millar}}, \bibinfo {author} {\bibfnamefont
  {G.}~\bibnamefont {Raffelt}}, \bibinfo {author} {\bibfnamefont
  {J.}~\bibnamefont {Redondo}}, \bibinfo {author} {\bibfnamefont
  {O.}~\bibnamefont {Reimann}}, \bibinfo {author} {\bibfnamefont
  {A.}~\bibnamefont {Ringwald}}, \bibinfo {author} {\bibfnamefont
  {K.}~\bibnamefont {Saikawa}}, \bibinfo {author} {\bibfnamefont
  {J.}~\bibnamefont {Schaffran}}, \bibinfo {author} {\bibfnamefont
  {A.}~\bibnamefont {Schmidt}}, \bibinfo {author} {\bibfnamefont
  {J.}~\bibnamefont {Sch{\"u}tte-Engel}}, \bibinfo {author} {\bibfnamefont
  {F.}~\bibnamefont {Steffen}}, \bibinfo {author} {\bibfnamefont
  {C.}~\bibnamefont {Strandhagen}}, \bibinfo {author} {\bibfnamefont
  {G.}~\bibnamefont {Wieching}},\ and\ \bibinfo {author} {\bibfnamefont {M.~A.
  D. M. A.~X.}\ \bibnamefont {Collaboration}},\ }\bibfield  {title} {\bibinfo
  {title} {A new experimental approach to probe qcd axion dark matter in the
  mass range above 40 $\mu$ev},\ }\href
  {https://doi.org/10.1140/epjc/s10052-019-6683-x} {\bibfield  {journal}
  {\bibinfo  {journal} {The European Physical Journal C}\ }\textbf {\bibinfo
  {volume} {79}},\ \bibinfo {pages} {186} (\bibinfo {year} {2019})}\BibitemShut
  {NoStop}%
\bibitem [{\citenamefont {Caldwell}\ \emph {et~al.}(2017)\citenamefont
  {Caldwell}, \citenamefont {Dvali}, \citenamefont {Majorovits}, \citenamefont
  {Millar}, \citenamefont {Raffelt}, \citenamefont {Redondo}, \citenamefont
  {Reimann}, \citenamefont {Simon},\ and\ \citenamefont
  {Steffen}}]{PhysRevLett.118.091801}%
  \BibitemOpen
  \bibfield  {author} {\bibinfo {author} {\bibfnamefont {A.}~\bibnamefont
  {Caldwell}}, \bibinfo {author} {\bibfnamefont {G.}~\bibnamefont {Dvali}},
  \bibinfo {author} {\bibfnamefont {B.}~\bibnamefont {Majorovits}}, \bibinfo
  {author} {\bibfnamefont {A.}~\bibnamefont {Millar}}, \bibinfo {author}
  {\bibfnamefont {G.}~\bibnamefont {Raffelt}}, \bibinfo {author} {\bibfnamefont
  {J.}~\bibnamefont {Redondo}}, \bibinfo {author} {\bibfnamefont
  {O.}~\bibnamefont {Reimann}}, \bibinfo {author} {\bibfnamefont
  {F.}~\bibnamefont {Simon}},\ and\ \bibinfo {author} {\bibfnamefont
  {F.}~\bibnamefont {Steffen}} (\bibinfo {collaboration} {MADMAX Working
  Group}),\ }\bibfield  {title} {\bibinfo {title} {Dielectric haloscopes: A new
  way to detect axion dark matter},\ }\href
  {https://doi.org/10.1103/PhysRevLett.118.091801} {\bibfield  {journal}
  {\bibinfo  {journal} {Phys. Rev. Lett.}\ }\textbf {\bibinfo {volume} {118}},\
  \bibinfo {pages} {091801} (\bibinfo {year} {2017})}\BibitemShut {NoStop}%
\bibitem [{\citenamefont {Baryakhtar}\ \emph {et~al.}(2018)\citenamefont
  {Baryakhtar}, \citenamefont {Huang},\ and\ \citenamefont
  {Lasenby}}]{PhysRevD.98.035006}%
  \BibitemOpen
  \bibfield  {author} {\bibinfo {author} {\bibfnamefont {M.}~\bibnamefont
  {Baryakhtar}}, \bibinfo {author} {\bibfnamefont {J.}~\bibnamefont {Huang}},\
  and\ \bibinfo {author} {\bibfnamefont {R.}~\bibnamefont {Lasenby}},\
  }\bibfield  {title} {\bibinfo {title} {Axion and hidden photon dark matter
  detection with multilayer optical haloscopes},\ }\href
  {https://doi.org/10.1103/PhysRevD.98.035006} {\bibfield  {journal} {\bibinfo
  {journal} {Phys. Rev. D}\ }\textbf {\bibinfo {volume} {98}},\ \bibinfo
  {pages} {035006} (\bibinfo {year} {2018})}\BibitemShut {NoStop}%
\bibitem [{\citenamefont {Chiles}\ \emph {et~al.}(2022)\citenamefont {Chiles},
  \citenamefont {Charaev}, \citenamefont {Lasenby}, \citenamefont {Baryakhtar},
  \citenamefont {Huang}, \citenamefont {Roshko}, \citenamefont {Burton},
  \citenamefont {Colangelo}, \citenamefont {Van~Tilburg}, \citenamefont
  {Arvanitaki}, \citenamefont {Nam},\ and\ \citenamefont
  {Berggren}}]{PhysRevLett.128.231802}%
  \BibitemOpen
  \bibfield  {author} {\bibinfo {author} {\bibfnamefont {J.}~\bibnamefont
  {Chiles}}, \bibinfo {author} {\bibfnamefont {I.}~\bibnamefont {Charaev}},
  \bibinfo {author} {\bibfnamefont {R.}~\bibnamefont {Lasenby}}, \bibinfo
  {author} {\bibfnamefont {M.}~\bibnamefont {Baryakhtar}}, \bibinfo {author}
  {\bibfnamefont {J.}~\bibnamefont {Huang}}, \bibinfo {author} {\bibfnamefont
  {A.}~\bibnamefont {Roshko}}, \bibinfo {author} {\bibfnamefont
  {G.}~\bibnamefont {Burton}}, \bibinfo {author} {\bibfnamefont
  {M.}~\bibnamefont {Colangelo}}, \bibinfo {author} {\bibfnamefont
  {K.}~\bibnamefont {Van~Tilburg}}, \bibinfo {author} {\bibfnamefont
  {A.}~\bibnamefont {Arvanitaki}}, \bibinfo {author} {\bibfnamefont {S.~W.}\
  \bibnamefont {Nam}},\ and\ \bibinfo {author} {\bibfnamefont {K.~K.}\
  \bibnamefont {Berggren}},\ }\bibfield  {title} {\bibinfo {title} {New
  constraints on dark photon dark matter with superconducting nanowire
  detectors in an optical haloscope},\ }\href
  {https://doi.org/10.1103/PhysRevLett.128.231802} {\bibfield  {journal}
  {\bibinfo  {journal} {Phys. Rev. Lett.}\ }\textbf {\bibinfo {volume} {128}},\
  \bibinfo {pages} {231802} (\bibinfo {year} {2022})}\BibitemShut {NoStop}%
\bibitem [{\citenamefont {McAllister}\ \emph {et~al.}(2018)\citenamefont
  {McAllister}, \citenamefont {Flower}, \citenamefont {Tobar},\ and\
  \citenamefont {Tobar}}]{PhysRevApplied.9.014028}%
  \BibitemOpen
  \bibfield  {author} {\bibinfo {author} {\bibfnamefont {B.~T.}\ \bibnamefont
  {McAllister}}, \bibinfo {author} {\bibfnamefont {G.}~\bibnamefont {Flower}},
  \bibinfo {author} {\bibfnamefont {L.~E.}\ \bibnamefont {Tobar}},\ and\
  \bibinfo {author} {\bibfnamefont {M.~E.}\ \bibnamefont {Tobar}},\ }\bibfield
  {title} {\bibinfo {title} {Tunable supermode dielectric resonators for axion
  dark-matter haloscopes},\ }\href
  {https://doi.org/10.1103/PhysRevApplied.9.014028} {\bibfield  {journal}
  {\bibinfo  {journal} {Phys. Rev. Applied}\ }\textbf {\bibinfo {volume} {9}},\
  \bibinfo {pages} {014028} (\bibinfo {year} {2018})}\BibitemShut {NoStop}%
\bibitem [{\citenamefont {Quiskamp}\ \emph {et~al.}(2020)\citenamefont
  {Quiskamp}, \citenamefont {McAllister}, \citenamefont {Rybka},\ and\
  \citenamefont {Tobar}}]{PhysRevApplied.14.044051}%
  \BibitemOpen
  \bibfield  {author} {\bibinfo {author} {\bibfnamefont {A.~P.}\ \bibnamefont
  {Quiskamp}}, \bibinfo {author} {\bibfnamefont {B.~T.}\ \bibnamefont
  {McAllister}}, \bibinfo {author} {\bibfnamefont {G.}~\bibnamefont {Rybka}},\
  and\ \bibinfo {author} {\bibfnamefont {M.~E.}\ \bibnamefont {Tobar}},\
  }\bibfield  {title} {\bibinfo {title} {Dielectric-boosted sensitivity to
  cylindrical azimuthally varying transverse-magnetic resonant modes in an
  axion haloscope},\ }\href {https://doi.org/10.1103/PhysRevApplied.14.044051}
  {\bibfield  {journal} {\bibinfo  {journal} {Phys. Rev. Applied}\ }\textbf
  {\bibinfo {volume} {14}},\ \bibinfo {pages} {044051} (\bibinfo {year}
  {2020})}\BibitemShut {NoStop}%
\bibitem [{\citenamefont {Lawson}\ \emph {et~al.}(2019)\citenamefont {Lawson},
  \citenamefont {Millar}, \citenamefont {Pancaldi}, \citenamefont
  {Vitagliano},\ and\ \citenamefont {Wilczek}}]{PhysRevLett.123.141802}%
  \BibitemOpen
  \bibfield  {author} {\bibinfo {author} {\bibfnamefont {M.}~\bibnamefont
  {Lawson}}, \bibinfo {author} {\bibfnamefont {A.~J.}\ \bibnamefont {Millar}},
  \bibinfo {author} {\bibfnamefont {M.}~\bibnamefont {Pancaldi}}, \bibinfo
  {author} {\bibfnamefont {E.}~\bibnamefont {Vitagliano}},\ and\ \bibinfo
  {author} {\bibfnamefont {F.}~\bibnamefont {Wilczek}},\ }\bibfield  {title}
  {\bibinfo {title} {Tunable axion plasma haloscopes},\ }\href
  {https://doi.org/10.1103/PhysRevLett.123.141802} {\bibfield  {journal}
  {\bibinfo  {journal} {Phys. Rev. Lett.}\ }\textbf {\bibinfo {volume} {123}},\
  \bibinfo {pages} {141802} (\bibinfo {year} {2019})}\BibitemShut {NoStop}%
\bibitem [{lnf(2021)}]{lnf_0p314}%
  \BibitemOpen
  \href@noop {} {\emph {\bibinfo {title} {LNF-LNC0.3\_14A 6-20 s/n 2122Z
  Cryogenic Low Noise Amplifier}}},\ \bibinfo {organization} {Low Noise
  Factory} (\bibinfo {year} {2021})\BibitemShut {NoStop}%
\bibitem [{\citenamefont {Faruque}(2017)}]{Faruque2017}%
  \BibitemOpen
  \bibfield  {author} {\bibinfo {author} {\bibfnamefont {S.}~\bibnamefont
  {Faruque}},\ }\bibinfo {title} {Frequency modulation (fm)},\ in\ \href
  {https://doi.org/10.1007/978-3-319-41202-3_3} {\emph {\bibinfo {booktitle}
  {Radio Frequency Modulation Made Easy}}}\ (\bibinfo  {publisher} {Springer
  International Publishing},\ \bibinfo {address} {Cham},\ \bibinfo {year}
  {2017})\ pp.\ \bibinfo {pages} {33--44}\BibitemShut {NoStop}%
\end{thebibliography}%

\end{document}